\documentstyle{mn}
\def\go{
\mathrel{\raise.3ex\hbox{$>$}\mkern-14mu\lower0.6ex\hbox{$\sim$}}
}
\def\lo{
\mathrel{\raise.3ex\hbox{$<$}\mkern-14mu\lower0.6ex\hbox{$\sim$}}
}

\def\etal{{\it et al.\ }}

\def\etal{{\it et al.\ }}

\def\be{\begin{equation}}
\def\ee{\end{equation}}
\def\bea{\begin{eqnarray}}
\def\eea{\end{eqnarray}}

\def\etal{{\sl et al.\ }}

\def\hw2{{\hat W}^2}
\def\go{\mathrel{\raise.3ex\hbox{$>$}\mkern-14mu
             \lower0.6ex\hbox{$\sim$}}}
\def\lo{\mathrel{\raise.3ex\hbox{$<$}\mkern-14mu
             \lower0.6ex\hbox{$\sim$}}}
\def\ltorder{\lo}
\def\gtorder{\go}

\def\eps2{{\epsilon^2}}

\def\msun{\hbox{$\hbox{M}_{\odot}$}}

\def\pc3{{\rm \, pc^{-3}}}

\input epsf.sty
\def\temp{1.34}%
\let\tempp=\relax
\expandafter\ifx\csname psboxversion\endcsname\relax
\else
    \ifdim\temp cm>\psboxversion cm
    \else
      \let\temp=\psboxversion
      \let\tempp= 
    \fi
\fi
\tempp
\let\psboxversion=\temp
\catcode`\@=11
\def\psfortextures{
\def\PSspeci@l##1##2{%
\special{illustration ##1\space scaled ##2}%
}}%
\def\psfordvitops{
\def\PSspeci@l##1##2{%
\special{dvitops: import ##1\space \the\drawingwd \the\drawinght}%
}}%
\def\psfordvips{
\def\PSspeci@l##1##2{%
\d@my=0.1bp \d@mx=\drawingwd \divide\d@mx by\d@my
\includegraphics{##1\space}}}%
\def\psforoztex{
\def\PSspeci@l##1##2{%
\special{##1 \space
      ##2 1000 div dup scale
      \number-\psllx\space \number-\pslly\space translate
}}}%
\def\psfordvitps{
\def\psdimt@n@sp##1{\d@mx=##1\relax\edef\psn@sp{\number\d@mx}}
\def\PSspeci@l##1##2{%
\special{dvitps: Include0 "psfig.psr"}
\psdimt@n@sp{\drawingwd}
\special{dvitps: Literal "\psn@sp\space"}
\psdimt@n@sp{\drawinght}
\special{dvitps: Literal "\psn@sp\space"}
\psdimt@n@sp{\psllx bp}
\special{dvitps: Literal "\psn@sp\space"}
\psdimt@n@sp{\pslly bp}
\special{dvitps: Literal "\psn@sp\space"}
\psdimt@n@sp{\psurx bp}
\special{dvitps: Literal "\psn@sp\space"}
\psdimt@n@sp{\psury bp}
\special{dvitps: Literal "\psn@sp\space startTexFig\space"}
\special{dvitps: Include1 "##1"}
\special{dvitps: Literal "endTexFig\space"}
}}%
\def\psfordvialw{
\def\PSspeci@l##1##2{
\special{language "PostScript",
position = "bottom left",
literal "  \psllx\space \pslly\space translate
  ##2 1000 div dup scale
  -\psllx\space -\pslly\space translate",
include "##1"}
}}%
\def\psforptips{
\def\PSspeci@l##1##2{{
\d@mx=\psurx bp
\advance \d@mx by -\psllx bp
\divide \d@mx by 1000\multiply\d@mx by \xscale
\incm{\d@mx}
\let\tmpx\dimincm
\d@my=\psury bp
\advance \d@my by -\pslly bp
\divide \d@my by 1000\multiply\d@my by \xscale
\incm{\d@my}
\let\tmpy\dimincm
\d@mx=-\psllx bp
\divide \d@mx by 1000\multiply\d@mx by \xscale
\d@my=-\pslly bp
\divide \d@my by 1000\multiply\d@my by \xscale
\at(\d@mx;\d@my){\special{ps:##1 x=\tmpx, y=\tmpy}}
}}}%
\def\psonlyboxes{
\def\PSspeci@l##1##2{%
\at(0cm;0cm){\boxit{\vbox to\drawinght
  {\vss\hbox to\drawingwd{\at(0cm;0cm){\hbox{({\tt##1})}}\hss}}}}
}}%
\def\psloc@lerr#1{%
\let\savedPSspeci@l=\PSspeci@l%
\def\PSspeci@l##1##2{%
\at(0cm;0cm){\boxit{\vbox to\drawinght
  {\vss\hbox to\drawingwd{\at(0cm;0cm){\hbox{({\tt##1}) #1}}\hss}}}}
\let\PSspeci@l=\savedPSspeci@l
}}%
\newread\pst@mpin
\newdimen\drawinght\newdimen\drawingwd
\newdimen\psxoffset\newdimen\psyoffset
\newbox\drawingBox
\newcount\xscale \newcount\yscale \newdimen\pscm\pscm=1cm
\newdimen\d@mx \newdimen\d@my
\newdimen\pswdincr \newdimen\pshtincr
\let\ps@nnotation=\relax
{\catcode`\|=0 |catcode`|\=12 |catcode`|
|catcode`#=12 |catcode`*=14
|xdef|backslashother{\}*
|xdef|percentother{
|xdef|tildeother{~}*
|xdef|sharpother{#}*
}%
\def\R@moveMeaningHeader#1:->{}%
\def\uncatcode#1{%
\edef#1{\expandafter\R@moveMeaningHeader\meaning#1}}%
\def\execute#1{#1}
\def\psm@keother#1{\catcode`#112\relax}
\def\executeinspecs#1{%
\execute{\begingroup\let\do\psm@keother\dospecials\catcode`\^^M=9#1\endgroup}}%
\def\@mpty{}%
\def\matchexpin#1#2{
  \fi%
  \edef\tmpb{{#2}}%
  \expandafter\makem@tchtmp\tmpb%
  \edef\tmpa{#1}\edef\tmpb{#2}%
  \expandafter\expandafter\expandafter\m@tchtmp\expandafter\tmpa\tmpb\endm@tch%
  \if\match%
}%
\def\matchin#1#2{%
  \fi%
  \makem@tchtmp{#2}%
  \m@tchtmp#1#2\endm@tch%
  \if\match%
}%
\def\makem@tchtmp#1{\def\m@tchtmp##1#1##2\endm@tch{%
  \def\tmpa{##1}\def\tmpb{##2}\let\m@tchtmp=\relax%
  \ifx\tmpb\@mpty\def\match{YN}%
  \else\def\match{YY}\fi%
}}%
\def\incm#1{{\psxoffset=1cm\d@my=#1
 \d@mx=\d@my
  \divide\d@mx by \psxoffset
  \xdef\dimincm{\number\d@mx.}
  \advance\d@my by -\number\d@mx cm
  \multiply\d@my by 100
 \d@mx=\d@my
  \divide\d@mx by \psxoffset
  \edef\dimincm{\dimincm\number\d@mx}
  \advance\d@my by -\number\d@mx cm
  \multiply\d@my by 100
 \d@mx=\d@my
  \divide\d@mx by \psxoffset
  \xdef\dimincm{\dimincm\number\d@mx}
}}%
\newif\ifNotB@undingBox
\newhelp\PShelp{Proceed: you'll have a 5cm square blank box instead of
your graphics (Jean Orloff).}%
\def\s@tsize#1 #2 #3 #4\@ndsize{
  \def\psllx{#1}\def\pslly{#2}%
  \def\psurx{#3}\def\psury{#4}
  \ifx\psurx\@mpty\NotB@undingBoxtrue
  \else
    \drawinght=#4bp\advance\drawinght by-#2bp
    \drawingwd=#3bp\advance\drawingwd by-#1bp
  \fi
  }%
\def\sc@nBBline#1:#2\@ndBBline{\edef\p@rameter{#1}\edef\v@lue{#2}}%
\def\g@bblefirstblank#1#2:{\ifx#1 \else#1\fi#2}%
{\catcode`\%=12
\xdef\B@undingBox{
\def\ReadPSize#1{
 \readfilename#1\relax
 \let\PSfilename=\lastreadfilename
 \openin\pst@mpin=#1\relax
 \ifeof\pst@mpin \errhelp=\PShelp
   \errmessage{I haven't found your postscript file (\PSfilename)}%
   \psloc@lerr{was not found}%
   \s@tsize 0 0 142 142\@ndsize
   \closein\pst@mpin
 \else
   \if\matchexpin{\GlobalInputList}{, \lastreadfilename}%
   \else\xdef\GlobalInputList{\GlobalInputList, \lastreadfilename}%
     \immediate\write\psbj@inaux{\lastreadfilename,}%
   \fi%
   \loop
     \executeinspecs{\catcode`\ =10\global\read\pst@mpin to\n@xtline}%
     \ifeof\pst@mpin
       \errhelp=\PShelp
       \errmessage{(\PSfilename) is not an Encapsulated PostScript File:
           I could not find any \B@undingBox: line.}%
       \edef\v@lue{0 0 142 142:}%
       \psloc@lerr{is not an EPSFile}%
       \NotB@undingBoxfalse
     \else
       \expandafter\sc@nBBline\n@xtline:\@ndBBline
       \ifx\p@rameter\B@undingBox\NotB@undingBoxfalse
         \edef\t@mp{%
           \expandafter\g@bblefirstblank\v@lue\space\space\space}%
         \expandafter\s@tsize\t@mp\@ndsize
       \else\NotB@undingBoxtrue
       \fi
     \fi
   \ifNotB@undingBox\repeat
   \closein\pst@mpin
 \fi
\message{#1}%
}%
\def\psboxto(#1;#2)#3{\vbox{%
   \ReadPSize{#3}%
   \advance\pswdincr by \drawingwd
   \advance\pshtincr by \drawinght
   \divide\pswdincr by 1000
   \divide\pshtincr by 1000
   \d@mx=#1
   \ifdim\d@mx=0pt\xscale=1000
         \else \xscale=\d@mx \divide \xscale by \pswdincr\fi
   \d@my=#2
   \ifdim\d@my=0pt\yscale=1000
         \else \yscale=\d@my \divide \yscale by \pshtincr\fi
   \ifnum\yscale=1000
         \else\ifnum\xscale=1000\xscale=\yscale
                    \else\ifnum\yscale<\xscale\xscale=\yscale\fi
              \fi
   \fi
   \divide\drawingwd by1000 \multiply\drawingwd by\xscale
   \divide\drawinght by1000 \multiply\drawinght by\xscale
   \divide\psxoffset by1000 \multiply\psxoffset by\xscale
   \divide\psyoffset by1000 \multiply\psyoffset by\xscale
   \global\divide\pscm by 1000
   \global\multiply\pscm by\xscale
   \multiply\pswdincr by\xscale \multiply\pshtincr by\xscale
   \ifdim\d@mx=0pt\d@mx=\pswdincr\fi
   \ifdim\d@my=0pt\d@my=\pshtincr\fi
   \message{scaled \the\xscale}%
 \hbox to\d@mx{\hss\vbox to\d@my{\vss
   \global\setbox\drawingBox=\hbox to 0pt{\kern\psxoffset\vbox to 0pt{%
      \kern-\psyoffset
      \PSspeci@l{\PSfilename}{\the\xscale}%
      \vss}\hss\ps@nnotation}%
   \global\wd\drawingBox=\the\pswdincr
   \global\ht\drawingBox=\the\pshtincr
   \global\drawingwd=\pswdincr
   \global\drawinght=\pshtincr
   \baselineskip=0pt
   \copy\drawingBox
 \vss}\hss}%
  \global\psxoffset=0pt
  \global\psyoffset=0pt
  \global\pswdincr=0pt
  \global\pshtincr=0pt 
  \global\pscm=1cm 
}}%
\def\psboxscaled#1#2{\vbox{%
  \ReadPSize{#2}%
  \xscale=#1
  \message{scaled \the\xscale}%
  \divide\pswdincr by 1000 \multiply\pswdincr by \xscale
  \divide\pshtincr by 1000 \multiply\pshtincr by \xscale
  \divide\psxoffset by1000 \multiply\psxoffset by\xscale
  \divide\psyoffset by1000 \multiply\psyoffset by\xscale
  \divide\drawingwd by1000 \multiply\drawingwd by\xscale
  \divide\drawinght by1000 \multiply\drawinght by\xscale
  \global\divide\pscm by 1000
  \global\multiply\pscm by\xscale
  \global\setbox\drawingBox=\hbox to 0pt{\kern\psxoffset\vbox to 0pt{%
     \kern-\psyoffset
     \PSspeci@l{\PSfilename}{\the\xscale}%
     \vss}\hss\ps@nnotation}%
  \advance\pswdincr by \drawingwd
  \advance\pshtincr by \drawinght
  \global\wd\drawingBox=\the\pswdincr
  \global\ht\drawingBox=\the\pshtincr
  \global\drawingwd=\pswdincr
  \global\drawinght=\pshtincr
  \baselineskip=0pt
  \copy\drawingBox
  \global\psxoffset=0pt
  \global\psyoffset=0pt
  \global\pswdincr=0pt
  \global\pshtincr=0pt 
  \global\pscm=1cm
}}%
\def\psbox#1{\psboxscaled{1000}{#1}}%
\newif\ifn@teof\n@teoftrue
\newif\ifc@ntrolline
\newif\ifmatch
\newread\j@insplitin
\newwrite\j@insplitout
\newwrite\psbj@inaux
\immediate\openout\psbj@inaux=psbjoin.aux
\immediate\write\psbj@inaux{\string\joinfiles}%
\immediate\write\psbj@inaux{\jobname,}%
\def\toother#1{\ifcat\relax#1\else\expandafter%
  \toother@ux\meaning#1\endtoother@ux\fi}%
\def\toother@ux#1 #2#3\endtoother@ux{\def\tmp{#3}%
  \ifx\tmp\@mpty\def\tmp{#2}\let\next=\relax%
  \else\def\next{\toother@ux#2#3\endtoother@ux}\fi%
\next}%
\let\readfilenamehook=\relax
\def\re@d{\expandafter\re@daux}
\def\re@daux{\futurelet\nextchar\stopre@dtest}%
\def\re@dnext{\xdef\lastreadfilename{\lastreadfilename\nextchar}%
  \afterassignment\re@d\let\nextchar}%
\def\stopre@d{\egroup\readfilenamehook}%
\def\stopre@dtest{%
  \ifcat\nextchar\relax\let\nextread\stopre@d
  \else
    \ifcat\nextchar\space\def\nextread{%
      \afterassignment\stopre@d\chardef\nextchar=`}%
    \else\let\nextread=\re@dnext
      \toother\nextchar
      \edef\nextchar{\tmp}%
    \fi
  \fi\nextread}%
\def\readfilename{\bgroup%
  \let\\=\backslashother \let\%=\percentother \let\~=\tildeother
  \let\#=\sharpother \xdef\lastreadfilename{}%
  \re@d}%
\xdef\GlobalInputList{\jobname}%
\def\psnewinput{%
  \def\readfilenamehook{
    \if\matchexpin{\GlobalInputList}{, \lastreadfilename}%
    \else\xdef\GlobalInputList{\GlobalInputList, \lastreadfilename}%
      \immediate\write\psbj@inaux{\lastreadfilename,}%
    \fi%
    \ps@ldinput\lastreadfilename\relax%
    \let\readfilenamehook=\relax%
  }\readfilename%
}%
\expandafter\ifx\csname @@input\endcsname\relax    
  \immediate\let\ps@ldinput=\input\def\input{\psnewinput}%
\else
  \immediate\let\ps@ldinput=\@@input
  \def\@@input{\psnewinput}%
\fi%
\def\nowarnopenout{%
 \def\warnopenout##1##2{%
   \readfilename##2\relax
   \message{\lastreadfilename}%
   \immediate\openout##1=\lastreadfilename\relax}}%
\def\warnopenout#1#2{%
 \readfilename#2\relax
 \def\t@mp{TrashMe,psbjoin.aux,psbjoint.tex,}\uncatcode\t@mp
 \if\matchexpin{\t@mp}{\lastreadfilename,}%
 \else
   \immediate\openin\pst@mpin=\lastreadfilename\relax
   \ifeof\pst@mpin
     \else
     \errhelp{If the content of this file is so precious to you, abort (ie
press x or e) and rename it before retrying.}%
     \errmessage{I'm just about to replace your file named \lastreadfilename}%
   \fi
   \immediate\closein\pst@mpin
 \fi
 \message{\lastreadfilename}%
 \immediate\openout#1=\lastreadfilename\relax}%
{\catcode`\%=12\catcode`\*=14
\gdef\splitfile#1{*
 \readfilename#1\relax
 \immediate\openin\j@insplitin=\lastreadfilename\relax
 \ifeof\j@insplitin
   \message{! I couldn't find and split \lastreadfilename!}*
 \else
   \immediate\openout\j@insplitout=TrashMe
   \message{< Splitting \lastreadfilename\space into}*
   \loop
     \ifeof\j@insplitin
       \immediate\closein\j@insplitin\n@teoffalse
     \else
       \n@teoftrue
       \executeinspecs{\global\read\j@insplitin to\spl@tinline\expandafter
         \ch@ckbeginnewfile\spl@tinline
       \ifc@ntrolline
       \else
         \toks0=\expandafter{\spl@tinline}*
         \immediate\write\j@insplitout{\the\toks0}*
       \fi
     \fi
   \ifn@teof\repeat
   \immediate\closeout\j@insplitout
 \fi\message{>}*
}*
\gdef\ch@ckbeginnewfile#1
 \def\t@mp{#1}*
 \ifx\@mpty\t@mp
   \def\t@mp{#3}*
   \ifx\@mpty\t@mp
     \global\c@ntrollinefalse
   \else
     \immediate\closeout\j@insplitout
     \warnopenout\j@insplitout{#2}*
     \global\c@ntrollinetrue
   \fi
 \else
   \global\c@ntrollinefalse
 \fi}*
\gdef\joinfiles#1\into#2{*
 \message{< Joining following files into}*
 \warnopenout\j@insplitout{#2}*
 \message{:}*
 {*
 \edef\w@##1{\immediate\write\j@insplitout{##1}}*
\w@{
\w@{
\w@{
\w@{
\w@{
\w@{
\w@{
\w@{
\w@{
\w@{
\w@{\string\input\space psbox.tex}*
\w@{\string\splitfile{\string\jobname}}*
\w@{\string\let\string\autojoin=\string\relax}*
}*
 \expandafter\tre@tfilelist#1, \endtre@t
 \immediate\closeout\j@insplitout
 \message{>}*
}*
\gdef\tre@tfilelist#1, #2\endtre@t{*
 \readfilename#1\relax
 \ifx\@mpty\lastreadfilename
 \else
   \immediate\openin\j@insplitin=\lastreadfilename\relax
   \ifeof\j@insplitin
     \errmessage{I couldn't find file \lastreadfilename}*
   \else
     \message{\lastreadfilename}*
     \immediate\write\j@insplitout{
     \executeinspecs{\global\read\j@insplitin to\oldj@ininline}*
     \loop
       \ifeof\j@insplitin\immediate\closein\j@insplitin\n@teoffalse
       \else\n@teoftrue
         \executeinspecs{\global\read\j@insplitin to\j@ininline}*
         \toks0=\expandafter{\oldj@ininline}*
         \let\oldj@ininline=\j@ininline
         \immediate\write\j@insplitout{\the\toks0}*
       \fi
     \ifn@teof
     \repeat
   \immediate\closein\j@insplitin
   \fi
   \tre@tfilelist#2, \endtre@t
 \fi}*
}%
\def\autojoin{%
 \immediate\write\psbj@inaux{\string\into{psbjoint.tex}}%
 \immediate\closeout\psbj@inaux
 \expandafter\joinfiles\GlobalInputList\into{psbjoint.tex}%
}%
\def\centinsert#1{\midinsert\line{\hss#1\hss}\endinsert}%
\def\psannotate#1#2{\vbox{%
  \def\ps@nnotation{#2\global\let\ps@nnotation=\relax}#1}}%
\def\pscaption#1#2{\vbox{%
   \setbox\drawingBox=#1
   \copy\drawingBox
   \vskip\baselineskip
   \vbox{\hsize=\wd\drawingBox\setbox0=\hbox{#2}%
     \ifdim\wd0>\hsize
       \noindent\unhbox0\tolerance=5000
    \else\centerline{\box0}%
    \fi
}}}%
\def\at(#1;#2)#3{\setbox0=\hbox{#3}\ht0=0pt\dp0=0pt
  \rlap{\kern#1\vbox to0pt{\kern-#2\box0\vss}}}%
\newdimen\gridht \newdimen\gridwd
\def\gridfill(#1;#2){%
  \setbox0=\hbox to 1\pscm
  {\vrule height1\pscm width.4pt\leaders\hrule\hfill}%
  \gridht=#1
  \divide\gridht by \ht0
  \multiply\gridht by \ht0
  \gridwd=#2
  \divide\gridwd by \wd0
  \multiply\gridwd by \wd0
  \advance \gridwd by \wd0
  \vbox to \gridht{\leaders\hbox to\gridwd{\leaders\box0\hfill}\vfill}}%
\def\fillinggrid{\at(0cm;0cm){\vbox{%
  \gridfill(\drawinght;\drawingwd)}}}%
\def\textleftof#1:{%
  \setbox1=#1
  \setbox0=\vbox\bgroup
    \advance\hsize by -\wd1 \advance\hsize by -2em}%
\def\textrightof#1:{%
  \setbox0=#1
  \setbox1=\vbox\bgroup
    \advance\hsize by -\wd0 \advance\hsize by -2em}%
\def\endtext{%
  \egroup
  \hbox to \hsize{\valign{\vfil##\vfil\cr%
\box0\cr%
\noalign{\hss}\box1\cr}}}%
\def\frameit#1#2#3{\hbox{\vrule width#1\vbox{%
  \hrule height#1\vskip#2\hbox{\hskip#2\vbox{#3}\hskip#2}%
        \vskip#2\hrule height#1}\vrule width#1}}%
\def\boxit#1{\frameit{0.4pt}{0pt}{#1}}%
\catcode`\@=12 
 \psfordvips   

\title[Binary--Binary Scattering]
{Close Approach during Hard Binary--Binary Scattering} 
\author[Bacon, Sigurdsson and Davies]{D.\ Bacon$^1$, S.\ Sigurdsson$^{2,3}$ and M.\ B.\ 
Davies$^2$\\
$^1$Jesus College, Cambridge CB5 8BL\\
$^2$Institute of Astronomy, Madingley Road, Cambridge CB3 OHA\\
$^3$Author to whom inquiries should be directed.}

\begin{document}

\date{Received ** *** 1995; in original form 1995 *** **}

\label{firstpage}

\maketitle

\begin{abstract}
It is now clear that there is a substantial population
of primordial binaries in galactic globular clusters and
that binary interactions are a major influence on globular
cluster evolution. Collisional interactions involving stars in binaries
may provide a significant channel for the formation
of various stellar exotica, such as blue stragglers, X--ray binaries
and millisecond pulsars. We report on an extensive series of
numerical experiments of binary--binary scattering, analysing
the cross--section for close approach during interactions for
a range of hard binary parameters of interest in globular cluster cores.
We consider the implied rate for tidal interactions for different
globular clusters and compare our results with previous, complementary
estimates of stellar collision rates in globular clusters.
We find that the collision rate for binary--binary encounters
dominates in low density clusters if the binary fraction in the cluster
is larger than $0.2$ for wide main--sequence binaries. In dense
clusters binary--single interactions dominate the collision rate
and the core binary fraction must be $\ltorder 0.1$ per decade in semi--major axis
or too many collisions
take place compared to observations. The rates are consistent if
binaries with semi--major axes $\sim 100 AU$ are overabundant in low density clusters
or if breakup and ejection substantially lowers the binary fraction
in denser clusters.
Given reasonable assumptions about fractions of binaries
in the cores of low density clusters such as NGC~5053, we cannot account
for all the observed blue stragglers by stellar collisions during binary
encounters, suggesting a substantial fraction may be due to coalescence
of tight primordial binaries.

\end{abstract}


\section{Introduction}

As the evidence for the presence of primordial
binaries in globular clusters increases, it has become clear
that the contribution of binary--single star and binary--binary scattering to
stellar collisions and other stellar binary processes must
be significant, at least in some clusters (see reviews by Hut \etal 1992;
Livio 1995; also,
Sigurdsson \& Phinney 1995, Davies 1995, Davies \& Benz 1995,
Leonard 1989, Goodman \& Hut 1989).
Of particular importance are tidal encounters, or stellar collisions,
that occur during resonances
that develop during hard binary--single and 
binary--binary scatterings.
These may contribute significantly to the formation of blue stragglers
(Leonard 1989, Leonard \&\ Fahlman 1991, Leonard \& Linnell 1992),
X--ray binaries, MSPs, CVs
runaway stars and other exotica (Sigurdsson \& Phinney 1995, Davies 1995).

Binary--binary scattering may be particularly important in low-density
clusters where there may be a large number of primordial binaries
and products of binary interactions such as blue stragglers
(Hills 1975, Nemec \& Harris 1987, Nemec \&\ Cohen 1989,
Leonard 1989, Mateo \etal 1990, Bolte 1991, Hills 1992,
Leonard \& Linnell 1992, Bolte \etal 1993, Yan \& Mateo 1994).
Mass segregation effects in globular clusters will increase
the binary fraction in the core compared to the rest of the cluster.
Thus, even if the binary fraction in the whole cluster is low (say
$\sim 5$\%) the fraction in the core may be much higher (see, for
example, Leonard 1989, Hut \etal 1992, McMillan \& Hut 1994,
but note also Sigurdsson \& Phinney 1995).

Here we report the results of 100,000 numerical experiments of
hard binary--binary scatterings, for a range of binary parameters
appropriate to globular cluster interactions.
Other studies of binary--binary encounters have been carried out
(Mikkola 1983, 1984a,b, Hoffer 1983,
Leonard 1989, Leonard \&\ Fahlman 1991, 
McMillan \etal 1990, 1991, Hut \etal 1992, Hut 1995, Rasio \etal 1995).
Mikkola considered a range of hard and soft binary scatterings
looking at the final state and energy transfer, while Hoffer included
mostly soft binary encounters. Leonard's work overlaps with ours, but
does not present a systematic survey of cumulative cross--sections
as reported here, and our work should be considered complementary to
his. McMillan, Hut and Rasio have so far mostly reported studies of particular
sets of encounters or encounters in particular models of clusters
rather than surveys of cross--sections.
We present a set of cumulative cross--sections for close
approach during hard encounters for a range of mass ratios and semi--major
axis ratios.
We compute the relative event rate for the various possible outcomes
of the encounters.
We also present sample cross--sections
for the change in semi--major axis during flybys
and compare them
to the one seen in encounters between binaries and single stars.
We leave a detailed discussion of the subsequent
evolution of the systems produced in encounters, such as triple-star
systems, to a later paper.

The cross--sections for close approaches calculated here complement
previous hydrodynamical calculations of the outcome of stellar collisions and
strong tidal interactions in the context of hard binary encounters
(Davies \etal 1994, Davies \& Benz 1995, Goodman \& Hernquist 1991,
Sigurdsson \& Hernquist 1992).

\section{Method}

The initial conditions for the scatterings were set following
the method of Hut \& Bahcall (1983, see also Sigurdsson \& Phinney 1993).
With two binaries, we have additional parameters from the relative
phase of the second binary, the orientation of the plane of the second
binary and the second binary mass ratio, semi--major axis
and eccentricity. We drew the binary parameters
by Monte Carlo selection uniformly over the phase variables.
The relative velocity at infinity of the centres--of--mass of the two
binaries, $v_{\infty}$ was chosen uniformly on the interval allowed.
We refer to a set of encounters performed at fixed $a_i, M_i$ and range of
$v_{\infty}$ as a ``run''.
A discrete set of values for the
semi--major axis, $a_1, a_2$ and masses, $M_1, M_2, M_3, M_4$
was used for each run.
The binary eccentricities, $e_{1,2}$ were zero for all encounters
reported here; previous calculations indicate the cross--sections of
interest are not sensitive to the binaries' eccentricities.
We discuss the limitations of this assumption later in this paper.
For the runs discussed here the binary parameters used
are shown in Table 1.

\begin{table}
\caption{The properties of all the runs performed.}
\begin{tabular}{rrrrrrrr} \hline\hline
Run & \multicolumn{7}{c}{Variables} \\
&${ v_{\infty}/v_c}$ &$a_1$  &$a_2$ &$M_{1,3}$ &$M_{2,4}$  &$v_c^2$ &N \\ \hline
\noalign{\vspace{0.3cm}}
10d &1/8-1/4 &1.0 &1.0 &1.0 &1.0 &2 &4000 \\
11d &1/4-1/2 &1.0 &1.0 &1.0 &1.0 &  &4000  \\
12d &1/2-1   &1.0 &1.0 &1.0 &1.0 &  &4000  \\
13d &1/16-1/8   &1.0 &1.0 &1.0 &1.0 &  &4000  \\
20d &1/8-1/4 &1.0 &1.0 &1.0 &0.5 &4/3 &4000  \\
21d &1/4-1/2 &1.0 &1.0 &1.0 &0.5 &  &4000  \\
22d &1/2-1   &1.0 &1.0 &1.0 &0.5 &  &4000  \\
10r &1/8-1/4 &2.0 &0.5 &1.0 &1.0 &5/2 &4000 \\
11r &1/4-1/2 &2.0 &0.5 &1.0 &1.0 &  &4000  \\
12r &1/2-1   &2.0 &0.5 &1.0 &1.0 &  &4000  \\
20r &1/8-1/4 &2.0 &0.5 &1.0 &0.5 &5/3 &4000  \\
21r &1/4-1/2 &2.0 &0.5 &1.0 &0.5 &  &4000  \\
22r &1/2-1   &2.0 &0.5 &1.0 &0.5 &  &4000  \\
30r &1/8-1/4 &$\sqrt{2}$ &$1/\sqrt{2}$  &1.0 &1.0 &$3/\sqrt{2}$ &4000 \\
31r &1/4-1/2 &$\sqrt{2}$ &$1/\sqrt{2}$ &1.0 &1.0 &  &4000  \\
32r &1/2-1   &$\sqrt{2}$ &$1/\sqrt{2}$ &1.0 &1.0 &  &4000  \\
40r &1/8-1/4 &$\sqrt{2}$ &$1/\sqrt{2}$ &1.0 &0.5 &$2/\sqrt{2}$ &4000  \\
41r &1/4-1/2 &$\sqrt{2}$ &$1/\sqrt{2}$ &1.0 &0.5 &  &4000  \\
42r &1/2-1   &$\sqrt{2}$ &$1/\sqrt{2}$ &1.0 &0.5 &  &4000  \\
50r &1/8-1/4 &4.0 &0.25 &1.0 &1.0 &17/4 &4000 \\
51r &1/4-1/2 &4.0 &0.25 &1.0 &1.0 &  &4000  \\
52r &1/2-1   &4.0 &0.25 &1.0 &1.0 &  &4000  \\
60r &1/8-1/4 &4.0 &0.25 &1.0 &0.5 &17/6 &4000  \\
61r &1/4-1/2 &4.0 &0.25 &1.0 &0.5 &  &4000  \\
62r &1/2-1   &4.0 &0.25 &1.0 &0.5 &  &4000  \\
\noalign{\vspace{0.3cm}}
\hline
\end{tabular}

\medskip

\end{table}

The critical velocity, $v_c$, is the velocity for which
the total energy of the system in the centre--of--mass frame
is zero, is given by
\begin{equation}
v_c^2 = {G\over {\mu}} \biggl (  { {M_1M_2}\over {a_1} }+{ {M_3M_4}\over {a_2} } \biggr )
\end{equation}

\noindent where $\mu = (M_1+M_2)(M_3+M_4)/M_T$, $M_T = M_1 + M_2 + M_3 + M_4$,
is the binaries reduced mass, and $a_1, a_2$ are the semi--major axes of
the binaries containing masses $M_{1,2}, M_{3,4}$ respectively. Note that
for these simulations $M_1 = M_3$ and $M_2 = M_4$. As a convention we
order $a_1 \geq a_2$, $M_{1,3} \geq M_{2,4}$, and choose $G=1$.
The sampling in velocity
was uniform in $v_{\infty}/v_c$ over the range indicated for each
set of runs shown. 

We refer to encounters where $v_{\infty}/v_c \leq 1$ as ``hard'',
following the nomenclature established for binary--single scatterings.
Hard encounters are dominated by gravitational focusing. Treating the
binaries as point masses at each binary centre--of--mass, for an
impact parameter $b$, the pericentre, $p$ is given by

\begin{equation}
p = { {GM_T}\over {v_{\infty}^2} }\biggl ( \Bigl (1 + b^2  \bigl ({ {v_{\infty}^2}\over {GM_T} } \bigr )^2  \Bigr )^{1/2} - 1 \biggr ).
\end{equation}

\noindent  For $v_{\infty}/v_c \ll 1$, $p \approx b^2 v_{\infty}^2/2GM_T$.
The impact parameter for each scattering is uniform in $b^2$ to
some maximum impact parameter $b_{max}$. By extension of
Hut \& Bahcall's choice (1983) we set $b_{max} = Ca_1/v_{\infty} + Da_1$,
where $C= 5, D= 0.6$ for the set of runs reported here. For
$v_{\infty}/v_c \ll 1$,

\begin{equation}
p_{max} = { {C^2 a_1^2}\over {2GM_T} } \biggl ( 1 + { {D v_{\infty}}\over {C} } \biggr )^2 .
\end{equation}

\noindent Note for $a_1 \gg a_2$ the maximum pericentre approach is
large compared to $a_2$; this is necessary as the wider binary may be sensitive to
perturbations from the tighter binary at several $a_1$, while the
tighter binary will likely be only very weakly perturbed. Thus for
$a_1/a_2 \gg 1$ we have to sample
the scatterings to large impact parameter.
For $a_1 = 4a_2$ and $a_1 = 16a_2$ we carried out a separate
run with $C=4$ to check the cross--section for very close approaches
had converged and we were sampling the strong interactions adequately.
The runs with $C=5$ proved adequate and results from those are reported
here to provide a homogenous sample. The results from the smaller
impact parameter runs will be discussed in a later paper.

While the simulations are scale free, the choice of masses
and $v_{\infty}/v_c$ were made bearing in mind the physically
interesting range of velocities in globular clusters,
$v_{\infty} \sim  10 {\rm km\, s^{-1}}$ and $M_i = 0.5 - 1.5 \msun $.
Of particular interest are wide ($a_i \sim 1-100 \, AU$) binaries
containing main--sequence stars near the turnoff ($M_i \approx 0.7 \msun $),
neutron stars ($M_i \approx 1.4 \msun $) and white dwarfs
($M_i \sim 0.5-1.2 \msun $).

\subsection{Integration scheme}

Two integration schemes were used in the calculations:
a fourth order Runge--Kutta integration scheme with adaptive
step size and quality control (see Hut \& Bahcall 1983, Sigurdsson
\& Phinney 1993), and a Bulirsch--Stoer variable step integrator
with KS--chain regularisation (Aarseth 1984, Mikkola 1983, 1984a,b).
The Runge--Kutta scheme is simple to implement and provided
a direct comparison with previous binary--single scatterings.
However, for $a_1/a_2 \gtorder 2$ the step size necessary to prevent
secular drift in the total energy of the system becomes prohibitively
small, and the Bulirsch--Stoer regularised scheme is an
order of magnitude faster
in integration despite the higher cost per integration step.
The Bulirsch--Stoer regularised scheme is more complicated to implement
and we relied heavily on subroutines provided by Sverre Aarseth.
Typically the Runge--Kutta integration for $a_1 = a_2$ required
$\gtorder 10^5$ steps to resolve a hard encounter, while the
Bulirsch--Stoer regularised scheme typically required $\ltorder 10^4$ steps.

As the Bulirsch--Stoer scheme uses large integration steps, there
is concern that it may not accurately track the true close approach
pair separations.
We note that the cross--sections from the set of runs carried out with 
the Runge--Kutta integration scheme
agreed to within statistical error with the cross--sections calculated
by the Bulirsch--Stoer regularised integrator for the same sets
of parameters. We also varied the parameter for integration tolerance
by two orders of magnitude for one set of runs, forcing a smaller
integration step size, and checked that the cross--sections did not
change with the integration step size. 
Here we report a homogenous set of runs done only with the
Bulirsch--Stoer regularised scheme.

The code was run on a DEC 3000/400 alpha, and the total set of runs
required about 3 weeks of cpu time. During each encounter the true pairwise
separation between each pair of particles was monitored and the
minimum value of each was stored. In addition we stored the position
and velocity of all four particles at the moment of the single
closest pair approach. This was for future analysis of the properties of
the remnant system for particles deemed to have undergone a strong
tidal encounter of collision. As the simulations are scale free,
we chose not to pick a scale and allow tidal interactions to
occur during the encounter, rather we analyse the outcome
using a ``sticky particle
approximation'' with a variable scale picked for each
run after the runs are completed
(Sigurdsson \& Phinney 1993, Davies \etal 1994).
The analysis using the sticky particle approximation
will be discussed in a later paper (in preparation).
The position
and velocities saved at closest approach also permit
generation of initial conditions for SPH simulations
by time reversing the integration from the point of
closest approach to a suitable initial separation
(see Sigurdsson \& Hernquist 1992).

The integration was stopped when either a maximum number
of integration steps had been taken, 2,000,000 for Runge--Kutta
integration, 500,000 for Bulirsch--Stoer integration; or
when two of the pair separations, $r_{ij}$, exceeded some
critical value $R_m = 1.2\times R_{in}$ where 
$R_{in} = 30\times \max \{a_1,a_2\}$
was the initial binary separation.

\subsection{Possible Outcomes}

There are several possible outcomes of encounters between two binaries.
In a flyby the two binaries may remain intact, 
with the components remaining unchanged,
but the orbital parameters of each binary may be perturbed, possibly
strongly.
Alternatively an exchange encounter may occur,
where components of the two binaries
are exchanged, producing two new binaries. The trajectories of the
four stars during such an encounter are shown in Figure 1.
Another possibility is that one binary may be broken
up resulting in the ejection of two
single stars (as illustrated in Figure 2), 
or one star may be ejected leaving the three
remaining stars in a triple.
The formation of
a triple-star system is shown in Figure 3.
Since we only consider hard encounters where the
total centre--of--mass energy is negative, it is
not possible for both binaries to be broken up by an
encounter.

\begin{figure}
\psboxto(\hsize;0cm){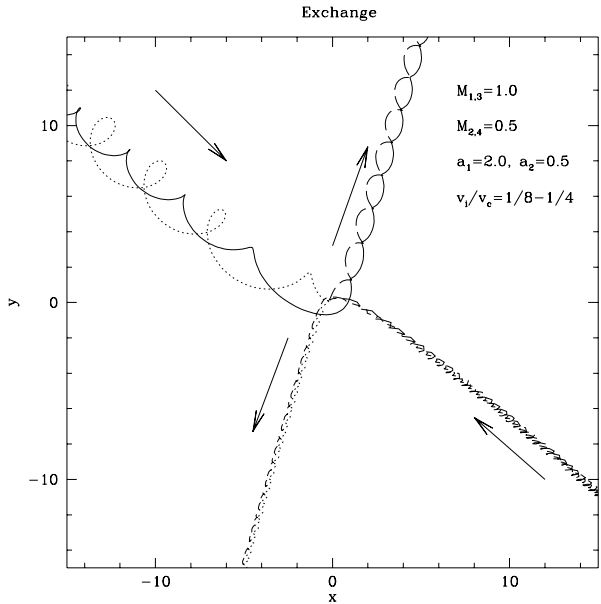}
\caption{An example of an exchange during an encounter
between unequal mass binaries with different semi--major axis.}
\end{figure}

\begin{figure}
\psboxto(\hsize;0cm){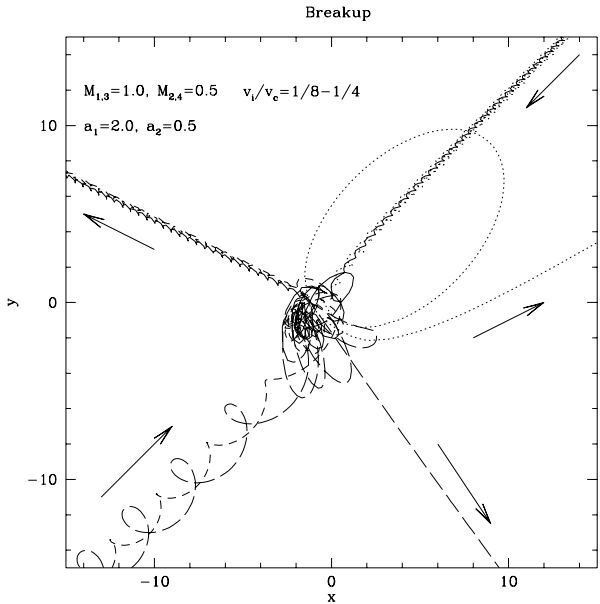}
\caption{An example of a breakup of a binary during a resonant encounter
between two binaries of unequal mass and semi--major axis.}
\end{figure}

\begin{figure}
\psboxto(\hsize;0cm){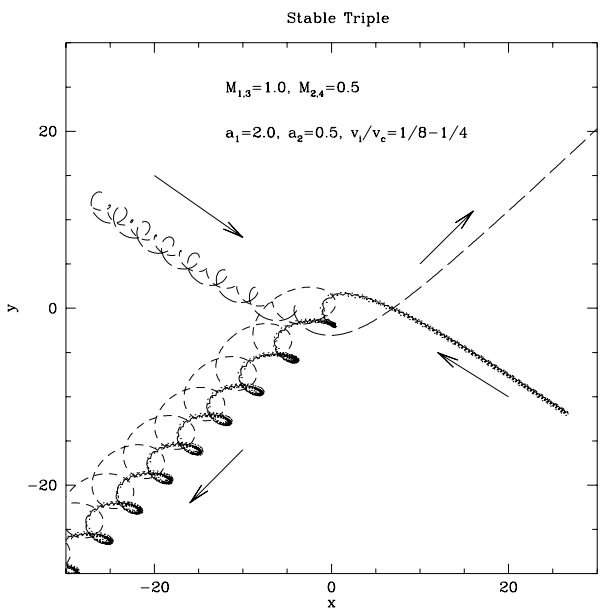}
\caption{A figure showing the ejection of a single star and the formation
of a triple-star system. As in previous figures the masses and initial semi--major
axis are unequal.}
\end{figure}

Table 2 shows the number of different outcomes for each of
the runs discussed here. As expected, the effectively larger
maximum pericentres for encounters with $a_1/a_2 \gg 1$
leads to proportionally more flybys in those runs.

\begin{table}
\caption{The frequency of the various outcomes.}
\begin{tabular}{rrrrrrr} \hline\hline
Run & \multicolumn{5}{c}{Outcome frequency} \\
& Flyby & Exchange & Breakup & Triple & Unresolved \\ \hline
\noalign{\vspace{0.3cm}}
10d & 2054 & 128 & 1256 & 429 & 133 \\
11d & 2290 & 127 & 1069 & 408 & 106 \\
12d & 2478 & 159 & 960 & 297 & 106 \\
13d & 1875 & 150 & 1255 & 514 & 206 \\
20d & 2534 & 98 & 871 & 371 & 126 \\
21d & 2702 & 107 & 832 & 288 & 71 \\
22d & 2821 & 101 & 820 & 181 & 77 \\
10r & 3485 & 21 & 196 & 250 & 48 \\
11r & 3344 & 21 & 298 & 294 & 43 \\
12r & 3437 & 16 & 398 & 138 & 11 \\
20r & 3434 & 14 & 221 & 260 & 71 \\
21r & 3522 & 13 & 202 & 236 & 27 \\
22r & 3561 & 11 & 301 & 116 & 11 \\
30r & 2867 & 50 & 673 & 308 & 112 \\
31r & 3051 & 56 & 587 & 233 & 73 \\
32r & 3184 & 50 & 564 & 140 & 62 \\
40r & 3143 & 36 & 507 & 242 & 74 \\
41r & 3268 & 34 & 448 & 208 & 42 \\
42r & 3328 & 36 & 454 & 142 & 40 \\
50r & 3669 & 6 & 51 & 225 & 49 \\
51r & 3679 & 2 & 156 & 162 & 1 \\
52r & 3674 & 3 & 305 & 16 & 2 \\
60r & 3755 & 0 & 31 & 186 & 28 \\
61r & 3754 & 1 & 116 & 127 & 2 \\
62r & 3761 & 2 & 220 & 17 & 0 \\
\noalign{\vspace{0.3cm}}
\hline
\end{tabular}

\medskip

\end{table}

The outcome of each encounter was analysed after the run was
complete to determine the final state of the binaries.
We do the final analysis of outcome after the runs are
terminated rather than ``on the fly''.
It is necessary to do the analysis after the runs,
because the termination conditions are not
exact and may lead to excessively long integrations for
individual encounters, to carry out the simulations in
a reasonably short time some fraction of unresolved
encounters must be accepted.
The pair of particles with the highest specific binding energy
was determined; this pair is assumed to be bound and its orbital
elements were solved for giving a new semi--major axis, $a_3$ and
eccentricity, $e_3$. As the total energy was negative, at least
one bound pair always exists. The other two particles were
then considered. If their separation from the most tightly bound pair
was less than $2a_3$ the stars were still considered to be
strongly interacting and thus the encounter was unresolved. If the
pair separation was larger, the orbital elements of the second
pair were solved for, neglecting any interaction with the first pair. If the resulting
eccentricity, $e_4$, was less than 1, we concluded two
bound binaries existed and we solved for the second semi--major axis, $a_4$.

If the separation of the two new binaries was increasing, and the
kinetic energy of the centres--of--masses of the binaries exceeded
the binary--binary binding energy, we assumed the binaries would
recede to infinity and we had a flyby or exchange, depending on
the membership of the respective binaries. Else we considered the
encounter unresolved.
Note that this is an approximate
condition, unlike the analogous case for three body scattering;
in practise the separations are sufficiently large to ensure
that few if any of these
cases were misclassified as resolved.

If the second pair of particles was not bound to each other, we considered the
particle furthest from the most tightly bound pair. If this particle
was receding from the remaining binary, and its kinetic energy exceeded
its binding energy relative to the other three particles, it
was considered to be escaping; if not, the encounter was
still in resonant interplay and was unresolved. Given that the furthest particle
was escaping, we now considered the remaining particle. We solved for
the orbital elements of the third particle about the new binary, treating
the binary as a point mass at its centre--of--mass. If the resulting
eccentricity, $e_t > 1$ the third star was unbound and we concluded 
that we had a binary and two single stars. If $e_t < 1$ the third
star was bound to the binary and we solved for its semi--major
axis relative to the binary centre--of--mass, $a_t$. To determine
whether the resulting triple was stable, we used the simple criterion,
$ 3\times a_3( 1 + e_3) < a_t (1 - e_t)$
(Harrington 1975, but see Kiselva \etal 1994).If this
criterion was satisfied, we assumed the triple was stable for timescales
much larger than the orbital timescales, while if it failed we
assumed the triple was unstable. Unstable triples are expected to
decay to a binary and a single star, with the new binary semi--major
axis drawn from a distribution approximately equal to that for
resonant binary--single scattering at the same total energy and
angular momentum. The approximate nature of 
the triple stability criterion 
is not a concern at this level of approximation; relatively few cases
are close to being marginally stable and if the onset of instability
is slow then for triples in globular clusters, encounters from
field stars are likely to determine the future dynamical evolution
of the system. A similar argument holds for binaries where the energy
transfer was to the binaries and the analysis suggests the stars will
not escape to infinity. For bound binary trajectories with apocentres $\gg a_{3,4}$, orbital
periods become long and in real clusters field star perturbations on
the quadruple start to become significant, the system is no longer dynamically
isolated and for us to continue the integration further becomes irrelevant physically.

\section{Dynamics and cross--sections}

For comparison with binary--single calculations and
in order to calculate physical interaction rates, it is
useful to calculate (normalised) cross--sections for
a process to take place during an encounter. This is
particularly important for comparing runs done with
different $a_{1,2}, M_i$ and with different $b_{max}$.

We define the cross--section for a process, $X$, to
be $\sigma_X = f(X)\pi b_{max}^2$, where $f(X)$ is the
fraction of the total encounters in that run where $X$ occurred
(Hut \& Bahcall 1983, Sigurdsson \& Phinney 1993).
We define a normalised cross--section, $\tilde\sigma_X$ by

\begin{equation}
\tilde\sigma_X = { {\sigma_X }\over {\pi (a_1^2 + a_2^2)} } 
\Bigl ( { {v_{\infty}}\over {v_c} } \Bigr )^2.
\end{equation}

\noindent This can be compared
with the similar definition for binary--single scattering
(Sigurdsson \& Phinney 1993). The normalisation to the binaries geometric
cross--section reduces to the binary--single case as $a_2 \rightarrow 0$
while the velocity ratio corrects for gravitational focusing. However,
because of the different scaling with $a_i$ for $p$, the resulting
rates are not as simple as for the binary--single case.

A quantity of particular interest is the rate at which a given process
$X$ occurs in a cluster (core) of density $n \pc3$ and velocity
dispersion $v_s$. If there are binaries (of the appropriate semi--major
axis) in the cluster (core), constituting some fraction $f_b$ of the
total number of stars, then we expect binaries to encounter each
other at some characteristic rate, $R_X$, with corresponding
mean time between $X$ occurring, $T_X = R_X^{-1}$. For a cluster with
1--D line--of--sight dispersion $v_s^{\prime }$,
we expect the mean binary encounter velocity
$ v_{\infty} \sim v_s \sim \sqrt{3}v_s^{\prime }$, as the binaries encounter
each other with the full 3--D relative velocity drawn from
the underlying velocity distribution, correcting for equipartition.
Deprojection of the low dispersion foreground and background contamination,
and the weighing of gravitational focusing provides additional
corrections of order $10\%$ (see Sigurdsson \& Phinney [1995] for discussion).

The rate for a process to take place for a given binary moving
in the cluster (core) is simply

\begin{equation}
R_X = \langle f_b n\sigma_X v \rangle .
\end{equation}

\noindent It is important in using this approximation to distinguish
the {\it global} binary fraction, that is the fraction of all stars
that are binaries, from the local binary fraction with the relevant
range of masses and semi--major axis. Relaxation
in a cluster may lead to a the fraction of binaries
with massive primaries in the core being higher than the
global binary fraction; while dynamical evolution will decrease
the core binary fraction (see Sigurdsson \& Phinney 1995, Hut \etal 1992).
A useful first approximation is to assume a global binary fraction
of $0.5$, with a uniform distribution in semi--major axis.
Then per decade in semi--major axis, 10\% of the stars are binaries
with a semi--major axis in that range.
Solving for $T_X$ substituting $\tilde\sigma_X$ and eliminating
$v_c^2$ we find

\begin{equation}
T_X = \Biggl < { { \langle {1 / {v_{\infty }}} \rangle^{-1} }\over 
{\pi f_b n (a_1^2+a_2^2) g(M) \Bigl (  {{M_1M_2}\over {a_1}} + 
{{M_3M_4}\over {a_2}} \Bigr )  } } { {1}\over {\tilde\sigma_X} } \Biggr > ,
\end{equation}

\noindent where $g(M) = GM_T/(M_1 + M_2)(M_3 + M_4)$.
It is useful to scale $T_X$ to the cluster parameters of
interest. Writing $n = 10^4 n_4 \pc3$,
$v_{\infty} = 10 v_{10}\ {\rm km\, s^{-1}}$ and defining
$a_{1,2} = a_{1,2}/AU$ and $M_i = M_i/\msun$, we can  write

\begin{equation}
T_X = {{1.5\times 10^{10}}\over { f_b g(M) n_4} } \Bigl ( (a_1^2+a_2^2)(E_{12} + E_{34}) \Bigl )^{-1}  \Bigl < {1\over{ v_{\infty } }} \Bigr >^{-1} {1\over {\tilde\sigma_X }} \, {\rm y},
\end{equation}

\noindent  where we have defined $E_{12} = M_1M_2/a_1$ and
$E_{34} = M_3M_4/a_2$ with the masses in solar masses and the
semi--major axis in $AU$ as before.

Table 3 shows the normalised cross--section for exchanges, breakups
and formation of stable triples for the different runs.

\begin{table}
\caption{Cross-sections for outcomes.}
\begin{tabular}{rrrrrrr} \hline\hline
Run & \multicolumn{3}{c}{$\tilde\sigma ({\rm outcome})$} \\
& Exchange & Breakup & Triple  \\ \hline
\noalign{\vspace{0.3cm}}
10d & 0.12 & 1.2 & 0.40  \\
11d & 0.23 & 2.0 & 0.75 \\
12d & 0.59 & 3.5 & 1.1  \\
13d & 0.069 & 0.58 & 0.24  \\
20d & 0.14 & 1.2 & 0.51  \\
21d & 0.30 & 2.3 & 0.80  \\
22d & 0.56 & 4.5 & 1.0  \\
10r & 0.014 & 0.14 & 0.17  \\
11r & 0.029 & 0.41 & 0.41  \\
12r & 0.044 & 1.10 & 0.38  \\
20r & 0.012 & 0.23 & 0.27  \\
21r & 0.027 & 0.42 & 0.49  \\
22r & 0.046 & 1.25 & 0.24  \\
30r & 0.049 & 0.66 & 0.30  \\
31r & 0.11 & 1.2 & 0.46  \\
32r & 0.20 & 2.2 & 0.55  \\
40r & 0.053 & 0.75 & 0.36  \\
41r & 0.10 & 1.3 & 0.61  \\
42r & 0.21 & 2.7 & 0.83  \\
50r & 0.0013  & 0.011 & 0.049  \\
51r & 0.00086 & 0.067 & 0.070 \\
52r & 0.0026 & 0.26 & 0.014 \\
60r & 0 & 0.010 & 0.060 \\
61r & 0.00065 & 0.075 & 0.082 \\
62r & 0.0026 & 0.28 & 0.022 \\
\noalign{\vspace{0.3cm}}
\hline
\end{tabular}

\medskip

\end{table}

An interesting result is the relatively low cross--section
for exchange, with a significantly higher cross--section
for formation of stable triples at all semi--major axis
ratios considered. As expected the relative cross--section
for breakup of one of the binaries is large and
increases both with $v_{\infty}/v_c$ and $a_2/a_1$. 
The fraction of unresolved encounters is small, except for the
hardest set of runs with $a_1 = a_2$. We don't expect the
unresolved encounters to contribute disproportionately to
the close approaches; they mostly consist of resonances
with one or two stars well separated from the other stars,
but with insufficient energy to reach infinity;
the other unresolved encounters consist of unstable triples.
We expect any ``memory'' of the initial conditions to have
been forgotten during the resonances, and the resonances are
resolved with a distribution exactly similar to the resonances
already resolved in that run, providing a few \% correction
to the cumulative cross--sections for close approach
at small $r_{min}$ and adding to the cross--sections for
different outcomes in the same proportion as the resolved encounters
(Heggie 1988, Sigurdsson \&\ Phinney 1993).
Note that for the high $a_1/a_2$ runs we find a significant
fraction of breakups of the wider binary despite the larger $b_{max}$
at constant $C$. With $a_1$ this large, these distant encounters
can still significantly perturb the wider binary.

\subsection{Close Approach}

A primary goal of this work is to estimate the rate
of tidal interactions during binary--binary encounters in
globular clusters, specifically encounters by hard binaries
in the cluster cores. In order to do that, it is desirable to
calculate the cumulative (normalised) cross--section for
close approach between any pair of stars during an encounter,
averaged over all of the binary phase space. Examples of these
cross--sections are shown in Figures 4--8, for different mass ratios,
semi--major axes and velocities. It is useful to fit the (normalised)
cross--sections with a simple broken power--law, as shown in
Figure 4 (see Hut \& Bahcall 1983, Sigurdsson \& Phinney 1993 for
comparable fits for binary--single scattering).

To produce a power law fit, we assume the cumulative cross--section,

\begin{equation}
\tilde\sigma ( \min \{ r_{ij} \} \leq r_{min} ) = \sigma_0 \Bigl ( {{r_{min}}\over a} \Bigr )^{\gamma } .
\end{equation}

\noindent In the case of binary--single scattering there is only
one length scale, $a$, to normalise the ``closeness'' of approach.
Here we have two length scales, $a_1, a_2$, and the normalisation
is more ambiguous. We find it useful to scale to the geometric
mean $a = \sqrt{a_1a_2}$ and this is what is plotted in Figures 4--8 and
is the normalisation used for calculating rates later. The choice
of $a$ is somewhat arbitrary for $a_1 \neq a_2$ and comparisons of
cross--section calculated by different authors must be done with care.

\begin{figure}
\psboxto(\hsize;0cm){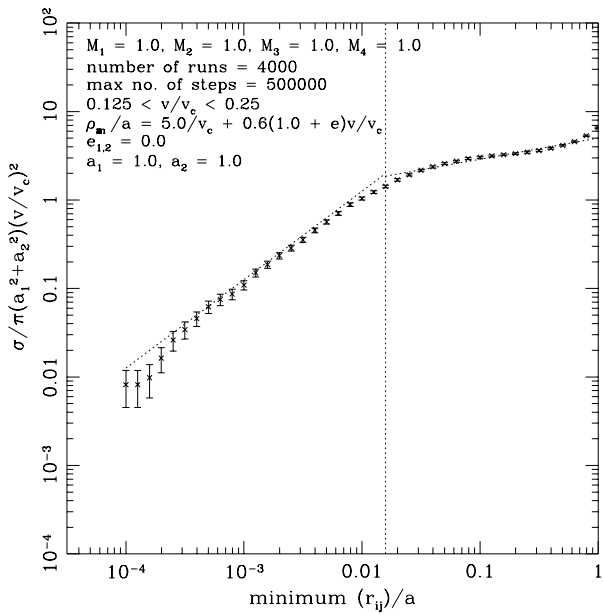}
\caption{
The cumulative cross--section for pairwise
close approach, $r_{ij}, i=1-4, j=i-4$, for particles $i,j$.
The plot shown is for equal
mass binaries with equal semi--major axis. The separation is scaled
to the geometric mean of the semi--major axis, $a = (a_1 a_2)^{1/2}$,
and the cross--section normalised to the geometric cross--section
of the binaries and corrected for gravitational focusing.
The solid lines show the piecewise power law fit to the cross--section,
with the dotted line showing the formal breaking point between the two
fits, $r_b$. The error bars are the standard error due to finite
sampling per separation bin.
}
\end{figure}

\begin{figure}
\psboxto(\hsize;0cm){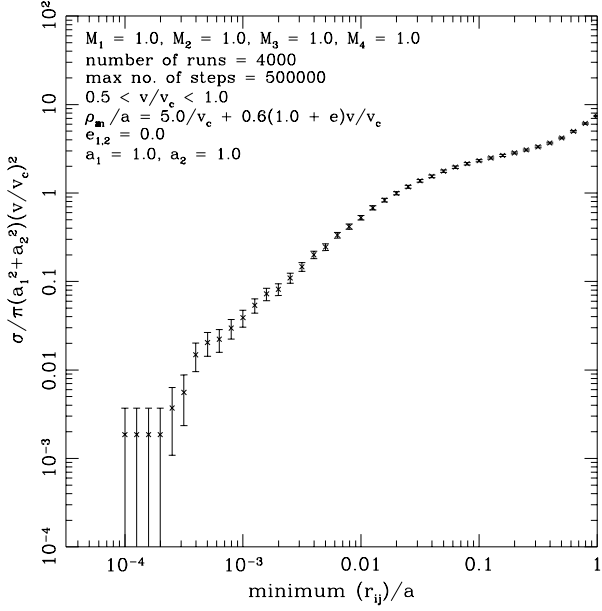}
\caption{The cumulative cross--section for close approach
for the high velocity run for equal mass and semi--major axis binaries.}
\end{figure}

\begin{figure}
\psboxto(\hsize;0cm){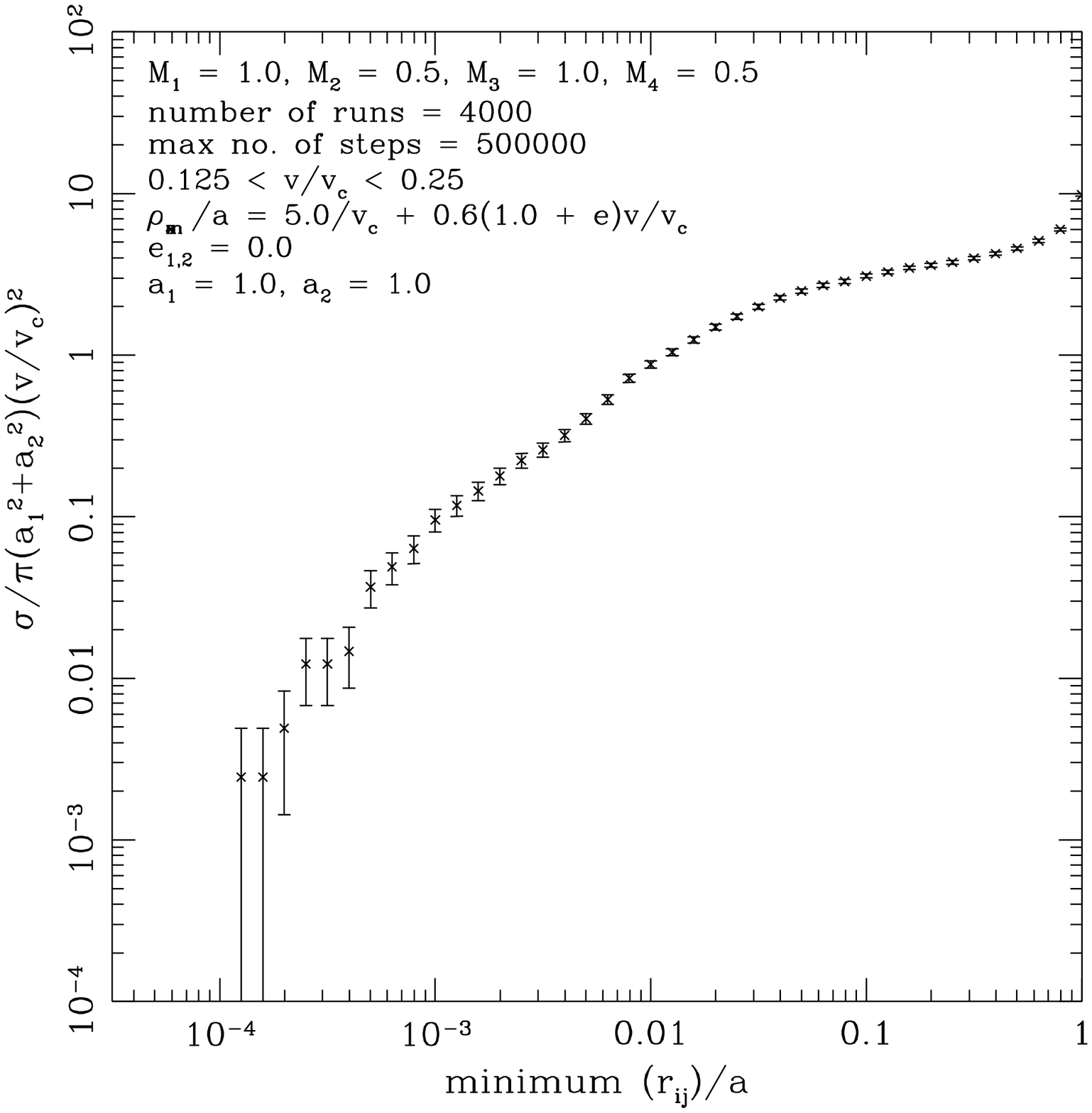}
\caption{The cumulative cross--section for close approach
for unequal mass and equal semi--major axis binaries.}
\end{figure}

\begin{figure}
\psboxto(\hsize;0cm){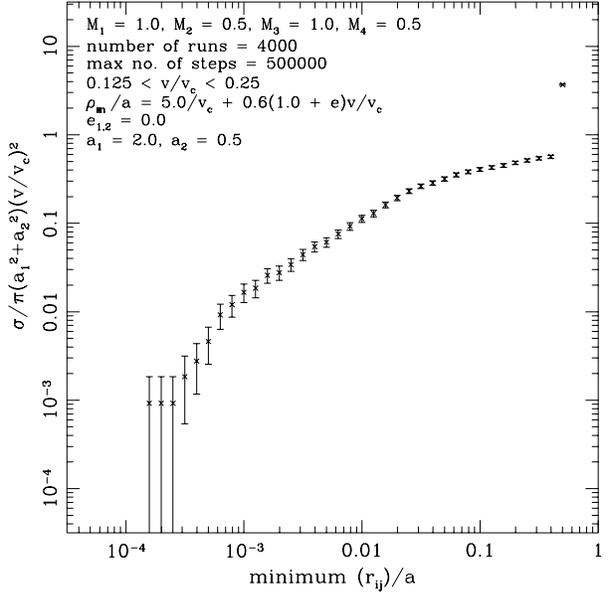}
\caption{The cumulative cross--section for close approach
for unequal mass and unequal semi--major axis binaries.
There is a sharp turnup in cross--section for $r_{ij} = a_2$,
this occurs because of the large number of encounters producing weak
perturbations to the tighter binary. The discrete binning in $r_{ij}$
does not resolve the turnup in cross--section.}
\end{figure}

\begin{figure}
\psboxto(\hsize;0cm){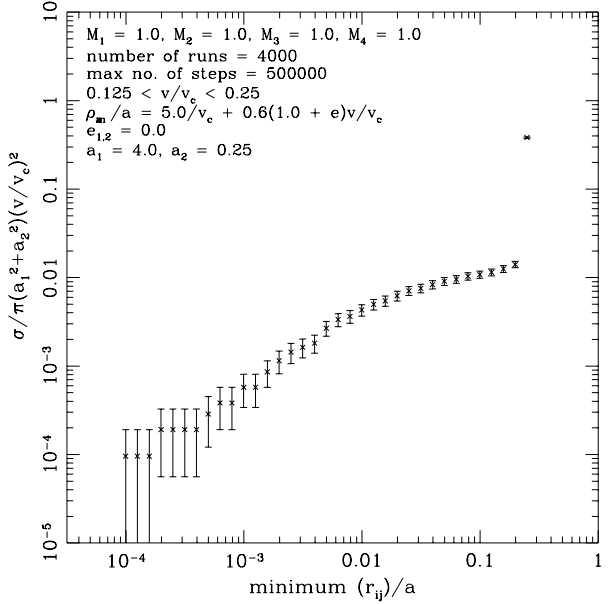}
\caption{The cumulative cross--section for close approach
for equal mass and unequal semi--major axis binaries.}
\end{figure}

To fit for $\sigma_0, \gamma $, we applied a piecewise (unweighted) least squares
fit to the log of the cumulative cross--section vs. $\log r_{ij}$.
The resulting fits are shown in Table 4. We define an additional
parameter $r_b$, the separation at which we switch from
one power law to the other. For $\min r_{ij} \leq r_b$,
$\sigma_0 = \tilde\sigma_1$
and $\gamma = \gamma_1$, whereas for
$\min r_{ij} > r_b$ $\tilde\sigma_2, \gamma_2$ should
be used respectively. $r_b$ is plotted as a dotted line in Figure 4 for
illustrative purposes. Formally $r_b$ is defined to be where the
fits cross; in practice this is close to the ``knee'' of the cumulative
cross--section curve as desired.
Because we oversample to large impact parameters, we discard the outermost
point ($r_{ij} = a_2$) in the cumulative cross--section when producing the
fit. It is necessary to use large enough impact parameters to be certain the
cross--section for close approaches has converged, but for our purposes we are
not interested in the weak perturbations to the semi--major axis caused by
the widest flybys.
For the high velocity runs with $a_1 = 16a_2$ a single power
law fit was adequate for all $r_{ij}$ as indicated in the table.

\begin{table}
\caption{The power law fits for $\tilde\sigma (r_{\rm min})$.}
\begin{tabular}{rrrrrr} \hline\hline
Run & $\tilde\sigma_1$ & $\tilde\sigma_2$ & $\gamma_1$ & $\gamma_2$ & $r_{\rm b}$ \\ \hline
\noalign{\vspace{0.3cm}}
10d & 130 & 5.0 & 1.0 & 0.24 & $1.6\times 10^{-2}$ \\
11d & 160 & 4.8 & 1.1 & 0.27 & $1.8\times 10^{-2}$ \\
12d & 210 & 5.7 & 1.3 & 0.41 & $1.7\times 10^{-2}$ \\
13d & 160 & 5.2 & 1.1  & 0.24  & $1.6\times 10^{-2}$ \\
20d & 150 & 6.0 & 1.1 & 0.32 & $1.8\times 10^{-2}$ \\
21d & 230 & 6.4 & 1.3 & 0.39 & $2.0\times 10^{-2}$ \\
22d & 120 & 6.8 & 1.2 & 0.46 & $2.0\times 10^{-2}$ \\
10r & 28 & 0.61 & 1.2 & 0.42 & $7.4\times 10^{-3}$ \\
11r & 9.5 & 0.73 & 0.98 & 0.39 & $1.3\times 10^{-2}$ \\
12r & 21 & 0.49 & 1.3 & 0.41 & $1.5\times 10^{-2}$ \\
20r & 5.8 & 0.9 & 0.86 & 0.35 & $2.6\times 10^{-2}$ \\
21r & 4.6 & 0.86 & 0.86 & 0.46 & $1.5\times 10^{-2}$ \\
22r & 4.0 & 0.61 & 0.98 & 0.49 & $2.2\times 10^{-2}$ \\
30r & 80 & 2.2 & 1.1 & 0.26 & $1.4\times 10^{-2}$ \\
31r & 40 & 2.1 & 1.0 & 0.31 & $1.4\times 10^{-2}$ \\
32r & 20 & 2.2 & 1.0 & 0.47 & $1.6\times 10^{-2}$ \\
40r & 49 & 2.7 & 1.1 & 0.35 & $2.1\times 10^{-2}$ \\
41r & 34 & 2.6 & 1.0 & 0.37 & $1.7\times 10^{-2}$ \\
42r & 96 & 2.8 & 1.3 & 0.49 & $1.3\times 10^{-2}$ \\
50r & 0.27 & 0.031 & 0.9 & 0.43 & $1.0\times 10^{-2}$ \\
51r & 15 & 0.019 & 1.8 & 0.42 & $8.0\times 10^{-3}$ \\
52r & 0.038 & - & 0.95 & - & - \\
60r & 0.41 & 0.027 & 1.1 & 0.43 & $1.7\times 10^{-2}$ \\
61r & 0.11 & 0.027 & 0.94 & 0.55 & $2.7\times 10^{-2}$ \\
62r & 0.048 & - & 0.89 & - & - \\
\noalign{\vspace{0.3cm}}
\hline
\end{tabular}

\medskip

\end{table}

The error--bars shown in Figures 4--8 show the standard deviation
due to sampling noise from the finite number of encounters per bin.
As the cross--section is cumulative the errors are systematically smaller
at larger $r_{ij}$, hence it is not appropriate to weight the least squares
fit by the errors.

As all the runs presented here are for $e_{1,2} = 0$ there might
be some concern that for binaries with a thermal eccentricity
distribution, as might be expected in globular clusters, the
cross--section for close approach could be substantially larger
than inferred from these simulations. To explore this we performed 
some runs with $e_{1,2}$ drawn from a thermal probability distribution,
$P(e_i) = 2e_i$, and compared the ratio of cumulative cross--section
for close approach for the eccentric and zero eccentricity binaries.

\begin{figure}
\psboxto(\hsize;0cm){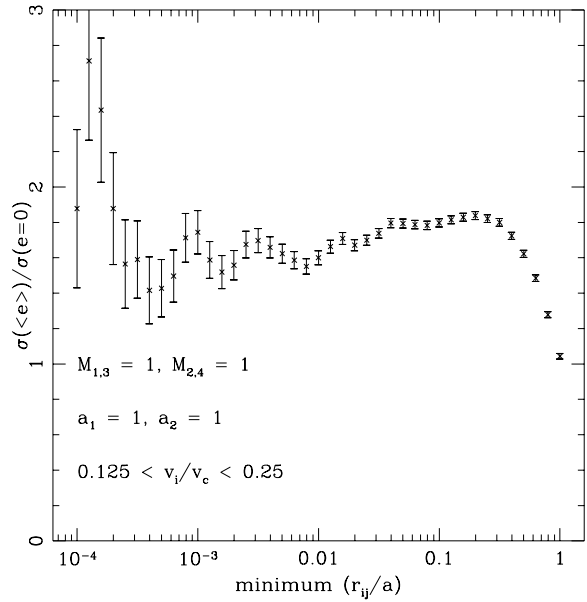}
\caption{The ratio of cumulative cross--section for close approach
for the for equal mass, equal semi--major axis binaries
for zero eccentricity and for $P(e) = 2e$ runs. The errorbars
show the standard error due to finite sampling.}
\end{figure}

The result is shown in Figure 9. As can be seen from the
figure, the ratio is approximately $1.7 = 1 + <e>$, over
a large range in $r_{ij}/a$. 
The ratio rises at small $r_{ij}/a$,
and results from other runs with $a_1 \neq a_2$
and $e_{1,2}$ thermal suggest this is significant.
The ratio approaches unity at $r_{ij}/a = 1$ by definition.
We conclude that allowing for an eccentricity distribution
increases the cross--section for collisions by a factor of $\sim 1.7$,
except for $ r_{ij}/a \sim 10^{-4}$ for which the correction
is somewhat larger. The same effect is seen for $a_2 \neq a_1$
but somewhat more pronounced. A possible concern is that this not
a real dynamical effect, but simply the high eccentricity end of
$a_2(1 - e_2)$ periastron passage. To check that this was not the case
we made two checks: we truncated the eccentricity distribution at $e_{1,2} = 0.98$
and checked the increase in relative cross--section was still present, as it was,
and we compared the initial periastron separation for each binary with $\min r_{ij}/a$.
The closest approaches were predominantly due to binaries with $e_{1,2} \sim 0.5$,
not encounters with very high initial binary eccentricity.

\subsection{Hardening of Binaries by Fly-by Encounters}

We now consider the effects on the binaries of fly-by
encounters. Such events are common compared to
other processes such as exchanges, or the formation of
triples as shown in Table 2. In Figures 10 and 12, we
plot the distribution of semi-major axes $a_1^{\prime}$, and $a_2^{\prime}$
of the two binaries after a fly-by encounter for $a_1 = a_2$ and
$a_1 = 4a_2$ respectively. As expected the distribution is
symmetric for the equal semi--major axis binaries. The empty region in
the upper right hand corner is forced by energy conservation, while
angular momentum conservation prevents both binaries from hardening
a lot simultaneously. 
For $a_1 = 4a_2$ most of the encounters produce a very weak perturbation
in $a_2$ compared with $a_1$, as expected.

\begin{figure}
\psboxto(\hsize;0cm){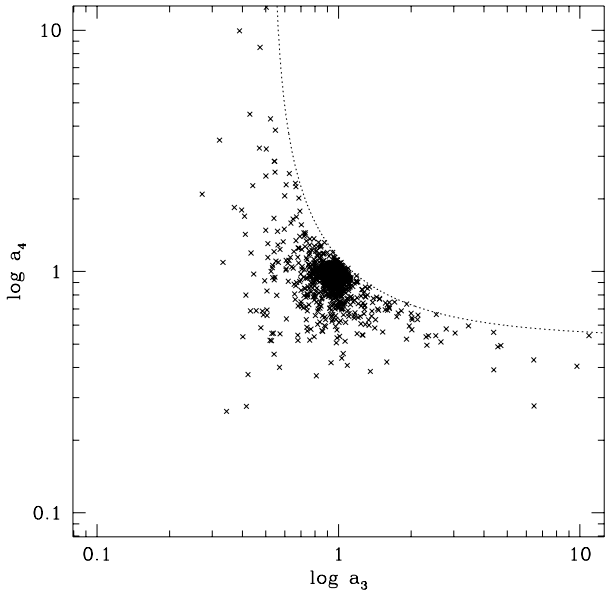}
\caption{
Plot of $\log a_3$ against $\log a_4$,
for flybys for equal mass and initial semi--major axis binaries.
There is a ``forbidden'' region excluded by
energy conservation, bounded by the dotted line in the figure.
The distribution in $a_3, a_4$ is roughly
symmetric as expected. The changes in semi--major axis included
775 encounters where $a_{3} < a_1$ and $a_2 < a_4$,
139 encounters where both binaries increased semi--major axis,
and 549 and 591 encounters where one binary increased its semi--major
axis and the other decreased, respectively.}
\end{figure}

\begin{figure}
\psboxto(\hsize;0cm){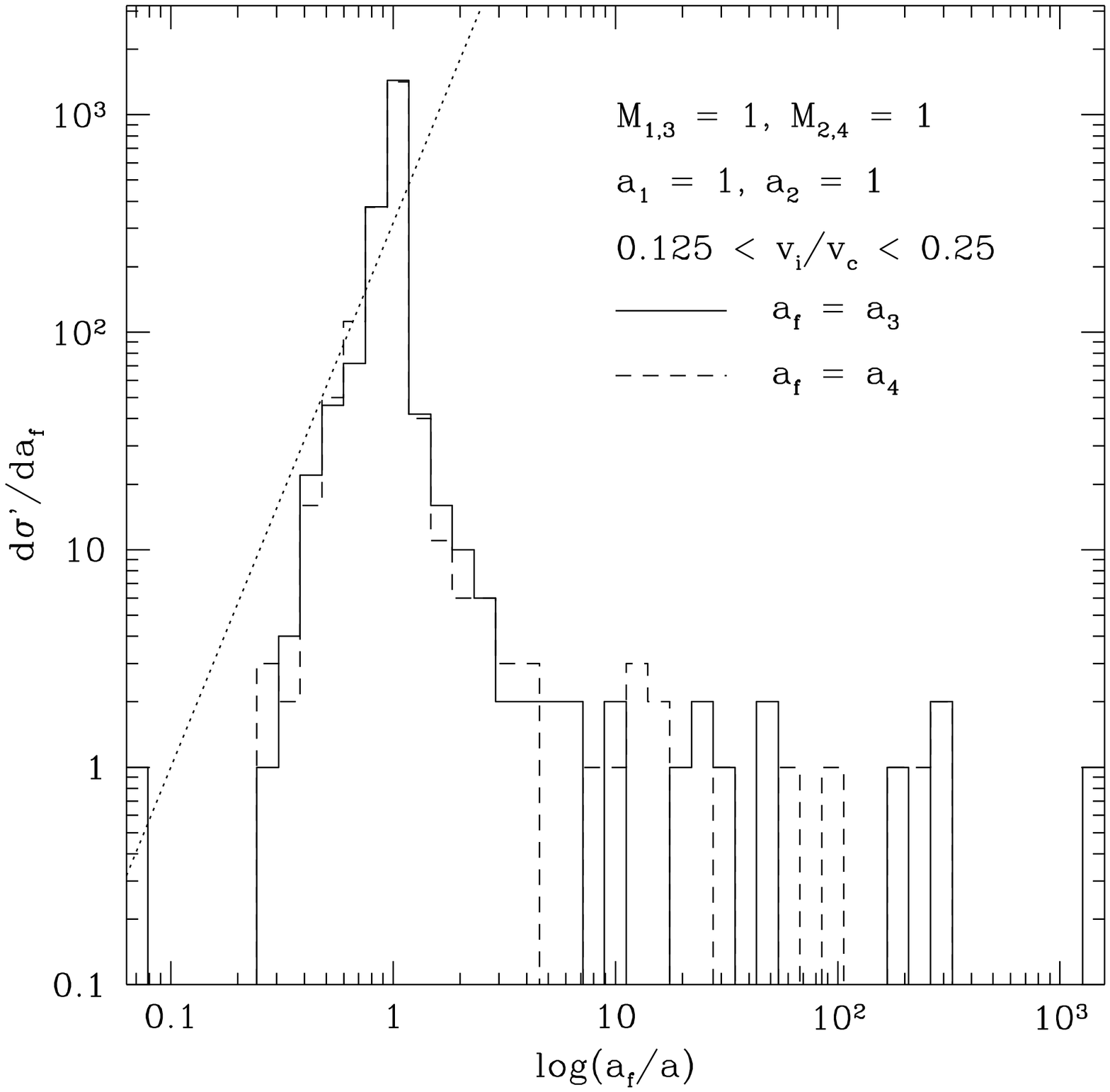}
\caption{Distribution of final semi--major axis. The solid histogram show
the distribution for $a_3$, the dashed line shows the distribution
for $a_4$. The cross-section is unnormalised. The dashed line shows
the distribution in $a_i$ expected from Heggie's law for binary--single
scattering, $d\sigma /da_i \propto a_i^{2.5}$.
The distribution shown in the figure is for equal mass
binaries and $a_1 = a_2$.
As expected, the distributions in $a_3, a_4$ are equal to within poisson noise.
The spike in the distribution at $a_{3,4} = 1$ is due to the large number
of wide encounters producing small changes in semi--major axis.
}
\end{figure}

\begin{figure}
\psboxto(\hsize;0cm){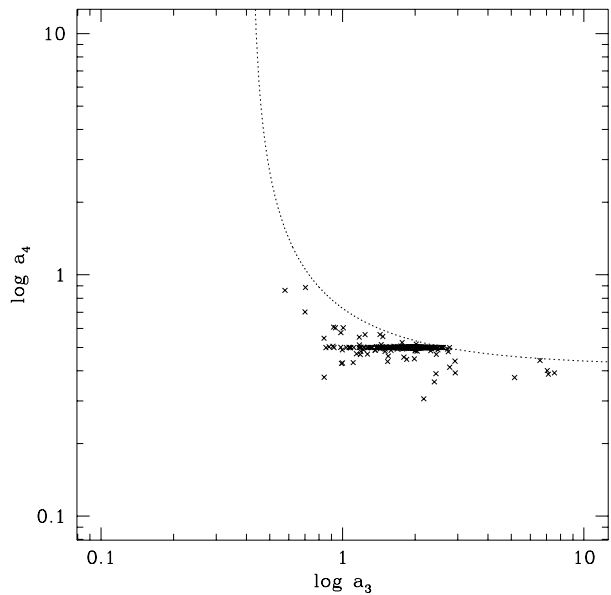}
\caption{Plot of $a_3$ against $a_4$,
for flybys for equal mass and unequal initial semi--major axis binaries.
The dotted line shows the ``forbidden'' region excluded by
energy conservation.
The changes in semi--major axis included
973 encounters where $a_{3} < a_1$ and $a_2 < a_4$,
770 encounters where both binaries increased semi--major axis,
648 encounters where $a_3 > a_1$ and $a_4 < a_2$, and
1094 encounters where $a_3 < a_1$ and $a_4 > a_3$.
}
\end{figure}

\begin{figure}
\psboxto(\hsize;0cm){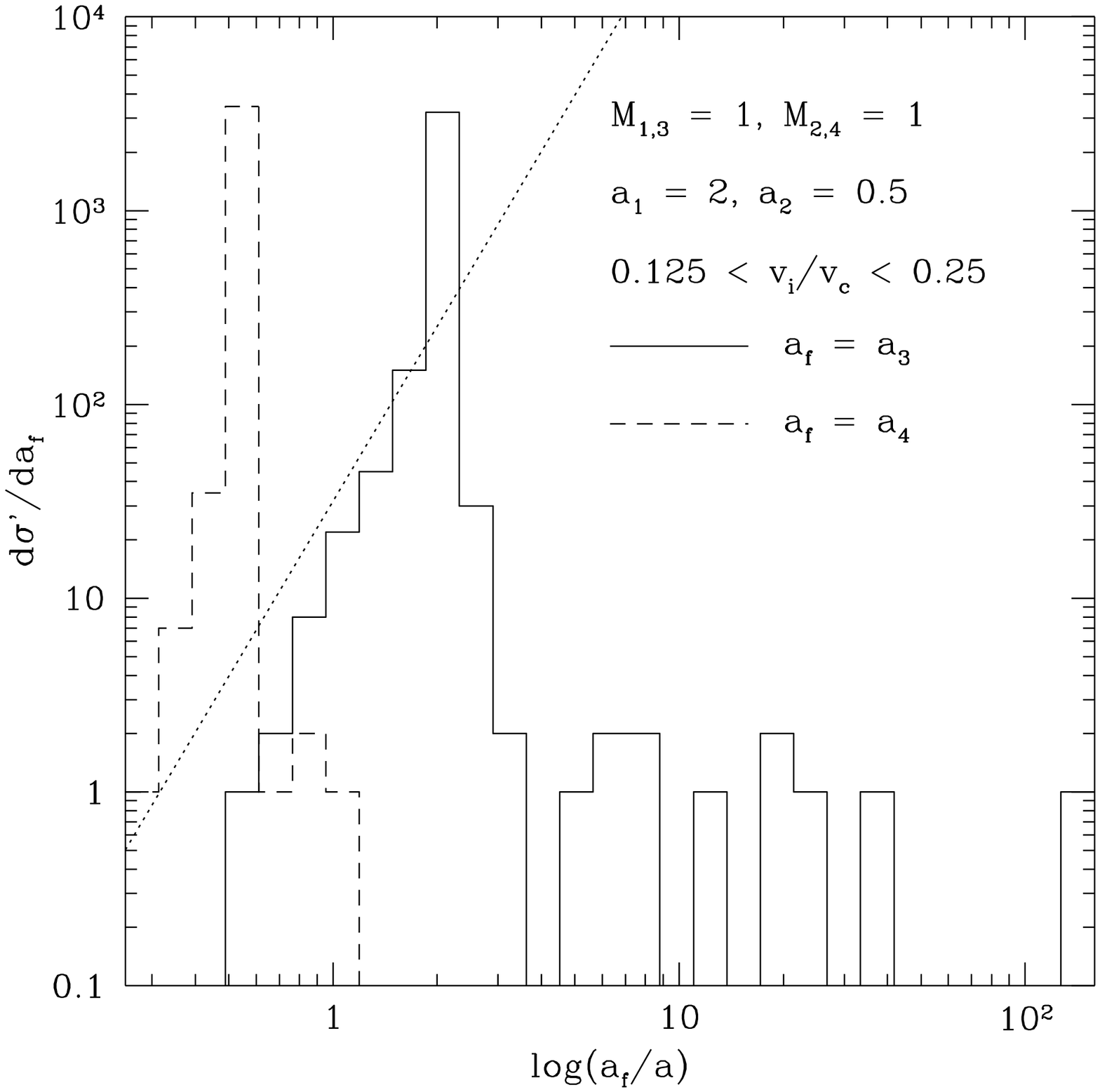}
\caption{Distribution of final semi--major axis. The solid histogram show
the distribution for $a_3$, the dashed line shows the distribution
for $a_4$. The cross-section is unormalised. The dashed line shows
the distribution in $a_i$ expected from Heggie's law for binary--single
scattering, $d\sigma /da_i \propto a_i^{2.5}$. 
The distribution shown here is for equal mass binaries and $a_1 = 4a_2$,
as expected the distributions in $a_3, a_4$ are different, $a_4$ showing
a much narrower spread from its initial value. Rather surprisingly, Heggie's
law still fits well for strong perturbations to the tighter binary,
for $a_4 \sim 1/2 a_2$.
The spike in the distribution at $a_{3} = 2, a_4 = 0.5$ is due to the large number
of wide encounters producing small changes in semi--major axis.
}
\end{figure}

Figures 11 and 13 show the differential cross--sections for change
in semi--major axis (cf. Davies \etal 1994, Sigurdsson \&\ Phinney 1993,
Hut \&\ Bahcall 1983). For binary--single scattering,
we expect $d\sigma/d\Delta \propto ( 1 + \Delta)^{-4.5}$ for strong
encounters (Heggie 1975), where $\Delta = \delta E_b/E_b$ is the fractional change
in binary binding energy. In semi--major axis space we expect
$d\sigma/da_i \propto a_i^{2.5}$ for strong encounters. The dotted
line in Figures 11 and 13 shows a $a_i^{2.5}$ power law, the differential
cross--section shows an approximate  fit to Heggie's law over an interesting range
in $\delta a/a$ for a sufficiently large $\delta a$. For weak perturbations,
the differential cross--section diverges with $b_{max}$ whereas
for very strong encounters the falloff in cross--section is somewhat
steeper than expected from Heggie's law.
It is interesting to note that resonant flybys do not lead
to large changes in binding energy compared to binary--single
encounters. For other classes of outcome we defer discussion
of the changes in energy to a later paper.

\subsection{Application of Cross--Sections to Compute Rates in 
Globular Clusters}

\begin{table}
\caption{The timescale for tidal interaction or stellar collision,
$T_C$ (in years), for a binary with
semi--major axis $a_1$ encountering a binary with semi--major axis $a_2$
in various globular cluster models, assuming $f_b = 1$ in the core for the
semi--major axis used. $a_1$ is shown in $AU$, and $M_\star$ is the mass of the
more massive star in each binary in solar masses. $v_s$ is given in ${\rm km\, s^{-1}}$.}
\begin{tabular}{rrrrrrr} \hline\hline
$n_\star $ &$ v_s $ &$a_1$ &${{v_{\infty}}\over {v_c}}$ 
&$M_\star$ &$a_1/a_2$ &$T_c$ \\ \hline
\noalign{\vspace{0.3cm}}
$10^2$ &$3$ &$100$  & 0.85 &0.7 &1 &$1.2\times 10^{12}$ \\
       & & &0.70  & &2 &$6.0\times 10^{11}$ \\
       & & &0.54 & &4 &$8.1\times 10^{10}$ \\
       & & &0.29 & &16 &$5.3\times 10^{11}$ \\
$10^2$ & &$100$  &0.74 &1.4 &1 &$6.0\times 10^{11}$ \\
       & & &0.60 & &2 &$1.3\times 10^{12}$ \\
       & & &0.46 & &4 &$3.0\times 10^{11}$ \\
       & & &0.25 & &16 &$7.7\times 10^{11}$ \\
$10^2$ &$10$ &$10$  &0.90 &0.7 &1 &$2.0\times 10^{12}$ \\
       & & &0.73 & &2 &$2.0\times 10^{12}$ \\
       & & &0.57 & &4 &$1.0\times 10^{12}$ \\
       & & &0.31 & &16 &$6.7\times 10^{12}$ \\
$10^2$ & &$10$  &0.78 &1.4 &1 &$1.3\times 10^{12}$ \\
       & & &0.63 & &2 &$2.2\times 10^{12}$ \\
       & & &0.49 & &4 &$8.1\times 10^{11}$ \\
       & & &0.27 & &16 &$7.2\times 10^{12}$ \\
$10^4$ &$5$ &$10$  &0.45 &0.7 &1 &$3.3\times 10^{9}$ \\
       & & &0.37 & &2 &$5.0\times 10^{9}$ \\
       & & &0.28 & &4 &$3.1\times 10^{9}$ \\
       & & &0.15 & &16 &$2.2\times 10^{10}$ \\
$10^4$ & &$10$  &0.39 &1.4 &1 &$6.8\times 10^{9}$ \\
       & & &0.32 & &2 &$4.4\times 10^{9}$ \\
       & & &0.25 & &4 &$1.5\times 10^{9}$ \\
       & & &0.13 & &16 &$1.9\times 10^{10}$ \\
$10^4$ &$10$  &$1$ &0.28 &0.7 &1 &$5.2\times 10^{9}$ \\
       & & &0.23 & &2 &$7.8\times 10^{9}$ \\
       & & &0.18 & &4 &$3.4\times 10^{10}$ \\
       & & &0.10 & &16 &$8.3\times 10^{10}$ \\
$10^4$ & &$1$  & 0.25 &1.4 &1 &$4.2\times 10^{9}$ \\
       & & &0.20 & &2 &$9.3\times 10^{9}$ \\
       & & &0.16 & &4 &$1.5\times 10^{10}$ \\
       & & &0.084 & &16 &$7.6\times 10^{10}$ \\
$10^6$ &$10$ &$0.3$ &0.16 &0.7 &1 &$8.0\times 10^{7}$ \\
       & & &0.13 & &2 &$1.9\times 10^{8}$ \\
       & & &0.10 & &4 &$1.2\times 10^{8}$ \\
       & & &0.053 & &16 &$9.4\times 10^{8}$ \\
$10^6$ & &$0.3$ &0.16 &1.4 &1 &$6.0\times 10^{7}$ \\
       & & &0.11 & &2 &$1.5\times 10^{8}$ \\
       & & &0.085 & &4 &$1.8\times 10^{8}$ \\
       & & &0.046 & &16 &$6.9\times 10^{8}$ \\
$10^6$ &$15$ &$0.1$ &0.13 &0.7 &1 &$2.8\times 10^{8}$ \\
       & & &0.11 & &2 &$6.5\times 10^{8}$ \\
       & & &0.085 & &4 &$1.5\times 10^{8}$ \\
       & & & -- & & & \\
$10^6$ & &$0.1$ &0.10 &1.4 & 1 &$2.1\times 10^{8}$ \\
       & & &0.095 & &2 &$4.7\times 10^{8}$ \\
       & & &0.074 & &4 &$3.1\times 10^{8}$ \\
       & & &0.04 & &16 &$9.3\times 10^{8}$ \\
\noalign{\vspace{0.3cm}}
\hline
\end{tabular}

\medskip

\end{table}

We now apply the cross--sections computed above to models of 
globular clusters. The calculated rates are listed with the
cluster parameters in Table 5.
Substituting equation 8 into equation 7, and approximating
$\langle 1/v_{\infty } \rangle ^{-1} = v_s $ we get
the rate of collisions

\begin{equation}
T_c = {{1.5\times 10^{10}}\over { f_b g(M) n_4 v_s} } \Bigl ( (a_1^2+a_2^2)(E_{12} + E_{34}) \Bigl )^{-1}
{{ (a_1 a_2)^{\gamma /2} }\over { \tilde\sigma_0 (r_{min})^{\gamma }  }} \, {\rm y},
\end{equation}

\noindent where $\gamma = \gamma_1, \tilde\sigma_0 = \tilde\sigma_1$ for $r_{min} \leq r_b$,
and $\gamma = \gamma_2, \tilde\sigma_0 = \tilde\sigma_2$ for $r_{min} > r_b$.
Table 5 shows a representative set of $T_C$ for a range of cluster parameters, and for
binaries containing two $0.7 \msun $ main--sequence stars and for binaries
containing a $1.4 \msun $ compact star and a $0.7 \msun $ main--sequence star
respectively. The binary separation assumed is shown for $a_1$ in $AU$, with
$a_1/a_2$ then as given in column 5.
For the calculation we choose $r_{min} = f_t R_*$, where
$R_*$ is the stellar radius, and $f_t \sim 3$ for main--sequence stars,
with $f_t R_*$ being the separation at which tidal effects become significant
in these encounters (Press \&\ Teukolsky 1977). As before, the masses are in solar
masses and radii in $AU$.
The timescales in Table 5 are calculated assuming $f_b = 1$, that
is all the stars in the cluster (core) are binaries with
the appropriate mass and semi--major axis. In practise $f_b < 1$, 
possibly much less, and the total collision rate represents an
average over all binary masses and semi--major axes.

A typical globular cluster core will contain about $N_*~\sim~10^4$ stars.
Here we follow observational conventions and count a binary
as a single ``star'' when figuring $N_*$. Approximating
the total collision rate as coming from a constant density core and neglecting
contributions due to binary interactions outside the core, an approximation
for the total number of collision products observed in a cluster is given by

\begin{equation}
N_c = f_e f_d f_b^2 N_* {{\tau }\over {T_c}},
\end{equation}

\noindent where $\tau $ is the characteristic lifetime over which
the collision product (eg. a blue straggler) is observable.
Here we separate the fractional binary density in the core,
$f_d$ and the binary fraction in the appropriate semi--major axis range, $f_b$.
We also allow for a factor $f_e$ to correct
for the eccentricity distribution. For most cases $f_e \sim 1.7$,
except noting that for $R_* \ltorder 10^{-3} a_2$
and $a_2 \ll a_1$, $f_e \sim 10$.
The true rate is the integrated rate over $T_c( M_i, a_i)$ given $f_b(M_i, a_i)$;
however the uncertainty in the binary population distribution is very large
and the systematic uncertainties in the integrand make a formal integral meaningless.
An approximation to the true collision rate can be made by assuming
that $O(6 \%)$ of the stars are near the turnoff, and another $O(2-4\%)$ is
in evolved remnants. Mass--segregation then increases the fractional density
of these more massive stars in the core further. As a simple approximation
we can also assume that the binary fraction, $f_b$, is about $0.1$ per decade in
semi--major axis independent of binary mass, and that the total initial
binary population spans about 5 decades in semi--major axis,
$a_i \sim 10^{-2} - 10^3 \, AU$.

In an old globular cluster, the core binary population is dynamically modified
by mass segregation and subsequent interactions.
Hardening and breakup reduce the number of wide binaries
but increase the fraction of binaries with semi--major axis, $\sim a_c$,
such that $T_R(a_c) \sim \tau_r$, where $\tau_r$ is the cluster
relaxation time and $T_R(a_c)$ is the characteristic time scale for a binary
with semi--major axis $a_c$ to change its semi--major axis by order $a_c/2$.
The situation is further complicated by dynamical recoil during strong
binary--single and binary--binary interactions, which tends to remove hardened
binaries from the cluster core to the cluster halo where interaction timescales
are long (see eg. Sigurdsson \& Phinney 1995). Binaries ejected to the cluster
halo then return to the cluster core on a relaxation timescale.

As can be seen from Table 5, the collision timescale for a binary of some semi--major
axis $a_1$ interacting with a binary of semi--major axis $a_2 \leq a_1$ does not
vary much from $a_1 = a_2$ to $a_1 \sim 10 a_2$, but then becomes longer
for $a_1 \gg 10 a_2$. Thus we can use a meaningful average collision timescale
for binaries with semi--major axis in the same decade span, and neglect collisional
interaction for binaries with semi--major axis in different decade spans as being
negligible by comparison. Consider three different clusters,
with core densities, $n$,
of $10^2, 10^4$ and $10^6 \pc3$ respectively. Assuming typical concentration
parameters of $W_0 = 6, 9, 12$ for multi--mass Michie--King models and a 
Salpeter initial mass--function, mass--segregation
increases the fractional density of $0.7+0.7 \msun $ binaries of all semi--major axes
from $0.06$, to $0.08, 0.15$ and $0.26$ respectively,
neglecting dynamical recoil.
Typical $\tau_r$ are $10^{10}, 10^9$ and $10^8$ years respectively, and we expect
those main--sequence binaries dominating the interaction rate to have semi--major axes of 
$10-100 AU$, $0.1-1 AU$ and $< 0.1 AU$ respectively. Assigning a binary fraction
of $f_b (a_i = 10-100 AU) \approx 0.2$ for the low density cluster,
$f_d = 0.08$ and $f_e = 10$, we find the expected
number of currently observable main--sequence collisions to be,
$N_{BS}(n = 10^2 \pc3) = 4 (f_b/0.2)^2$ per $10^4$ core stars
for $\tau_{BS} = 5\times 10^9$ years,
and taking $T_c \approx 10^{11}$ years as indicated in Table 5.
For the medium density cluster, we similarly find
$N_{BS}(n = 10^4 \pc3) = 40$ for
$f_b(a_i = 0.1-1 AU) = 0.2$, $f_d = 0.15$ and $f_e = 1.7$.
A higher $f_b$ may be appropriate as the interaction
time scale is short enough for a substantial fraction of binaries to be
hardened from larger semi--major axes to the optimum range for collisions.
For the densest cluster, we expect $f_d$ to be smaller than mass--segregation
would indicate as dynamical recoil and breakup are likely to have been significant
over the cluster history, so we take 
$f_e = 1.7$, $f_b(a_i \leq 0.1 AU) = 0.1$ and $f_d = 0.1$,
even though mass--segregation would suggest $f_d \sim 0.2-0.3$. Using these values
we find $N_{BS}(n = 10^6 \pc3) = 500$. However, the true number may be
smaller still as exchanges and collisions may have reduced the fraction of
core stars near the main--sequence turnoff relative to the evolved remnant
stars (see Sigurdsson \&\ Phinney 1995).

It is interesting to compare, for binaries in the same semi--major
axis range, the timescales to breakup due to binary--binary interactions
with the timescale to collision.
Using the results in Table 3,
and binary parameters as above,
we find $T_{breakup}( n = 10^2 \pc3 ) = 3\times 10^{10}$ years,
$T_{breakup}( n = 10^4 \pc3 ) = 4\times 10^{10}$ years and
$T_{breakup}( n = 10^6 \pc3 ) =  3\times 10^9$ years.
For the lowest density cluster, a binary is several times more likely to
be broken up by a binary--binary interaction, than be involved in
a stellar collision, whereas for the medium density cluster breakup
and collision are about equally likely, and the chance of a breakup
in a binary--binary collision in the densest cluster is somewhat
less than the probability of a stellar collision.


\subsection{Comparison with Previous Results
}

NGC~5053 is a very low density ($n = 8 \pc3 $), low dispersion globular,
with 24 observed candidate bue stragglers (Nemec \&\ Cohen 1989).
Some controversy exists in the literature over whether the blue stragglers
in this cluster must be due to merger through internal evolution of
initially tight primordial binaries (Hills \&\ Day 1976) or whether
binary--binary collisions may have produced a significant fraction
of the observed blue straggler stars (Leonard \&\ Fahlman 1991).

We find that we cannot produce the 24 blue stragglers observed in NGC 5053
even with $f_b =1$ for binaries with $a_1 \gtorder 100 AU$;
rather we find, $T_c \approx 2\times 10^{12}$ years, giving 
$N_{BS} = 2-3 (f_e/10)(f_d/0.1)(f_b/0.2)^2 $ for a total core
population of 40,000 stars. The encounter rate is dominated by
binaries with semi--major axes  $\gtorder 100 \, AU$, and it is
unlikely $f_b \gg 0.2$ for this semi--major axis range, as many binaries
must be much tighter than this. It is very unlikely that mass segregation
has increased the fraction of binaries with primaries of near turnoff
mass to $\gg 0.1$ in this cluster; the relaxation time scale is very long
and the concentration low ($c = 0.75$).

It is not necessary to produce
all the blue stragglers through binary--binary collisions, as some undoubtedly
formed through spiral--in of tight primordial binaries (see eg. Livio 1993).
It is possible that the fraction of primordial binaries with initial semi--major
axis small enough for them to merge in a globular cluster lifetime
is a function of the globular cluster initial density and dispersion.
It is also possible that the core density of low density globular clusters
has decreased on timescales of ${\rm few}\times 10^9$ years through
tidal shocking by the galaxy, and that the collision rates were thus
higher in the past. Note that for every blue straggler produced by
collisions during binary--binary encounters 
in these globular clusters we expect $\gtorder 10$ binaries
to have been dynamically broken up, and thus the current binary
fraction in the right semi--major axis range would be smaller
now than in the past.

In the
case of NGC 5053, we might require the core density to have been an
order of magnitude higher within the lifetime of the currently existent
blue stragglers, and a large fraction ($\gtorder 0.3$) of the core turnoff stars
to have been in wide ($a \sim 100 AU$) binaries. We do expect the binary fraction
to have decreased with time, but we still expect the present core binary fraction to
be $\sim 0.25$ if binary--binary collisions are to account for the blue stragglers.
This would be apparent in high precision photometry of NGC 5053 as a prominent second
main--sequence (Romani \&\ Weinberg 1991). Alternatively low density clusters
like NGC 5053 may form a significant number of wide triples, and collisions are
frequent during triple encounters.

It is instructive to compare the collision rate due to binary--binary encounters
with the collision rate due to binary--single encounters.
For the sample clusters discussed above, we find binaries
with those semi--major axes have collision timescales with
a single main--sequence star of $4.6\times 10^{11}$, $5.4\times 10^9$
and $1.8\times 10^8$ years respectively, and the expected number of
blue stragglers formed is given by $N_{BS} = f_df_b(1 - f_b) f_* N_* \tau /T_c$,
where $f_*$ is the fractional density of main--sequence stars at the turnoff
in the core, corrected for mass segregation ($0.13, 0.2, 0.2$ respectively).
Using the same $f_b$ as above, we find the expected number of blue stragglers
to be $0-1, 50$ and $1100$ respectively. That is, in the low density clusters
binary--single collisions rates are comparable to binary--binary collisions,
and the binary--singles dominate in the denser clusters. As not that many
blue stragglers are observed in globular clusters one might infer
the binary fraction per decade in semi--major axis is less
than the $0.1$ used here and the global binary fraction somewhat
less than $0.5$, which is consistent with observational estimates
(Pryor \etal 1989, Yan \&\ Mateo 1994) while still allowing a
sufficiently high collision rate to produce the blue stragglers and other
stellar exotica observed.

\section{Conclusions}

Some care must be taken in considering the effectiveness
of binary--binary collisions in globular clusters.
The {\it global} binary fraction at zero age in
clusters is probably $0.5-1.0$, comparable with that
seen in the field. However, this includes binaries
from near contact, $a_i \sim 0.01 AU$, to extremely
wide binaries, $a_i \gg 10^3 AU$. The former do
not interact on short enough timescales to be of interest,
except in core--collapsed clusters, and will in due course
merge through their internal evolution; the latter are soft
and have high encounter rates in all except the very lowest
density clusters, and are broken up in a few dynamical timescales.
The global binary fraction as a function of $a_i$ seems to
have an initial distribution of approximately $0.1$ per decade
in $a_i$, and that is the approximation we use above.
However, the core population of binaries is modified by
several processes, including mass--segregation, dynamical
recoil, exchange and breakup. As a result the fraction
of binaries in the core, per decade in $a_i$ and at different
masses varies with time and cluster parameters. In
calculating the expected number of blue stragglers above
we made some effort to correct for the dominant processes in
the different clusters considered.

We find collision timescales for plausible binary populations
comparable to the lifetime of the clusters, and an expected
number of blue stragglers sufficient to account for a large fraction
of the low density blue straggler population, but overestimating
the population in the denser clusters. This can be understood in
terms of the dynamical evolution of the globular cluster binary
population, as breakup and ejection decreases the core
population of binaries.

We have refrained here from discussing in detail the properties
of the final state of the binaries. In particular, parameters of
interest include the final distribution of semi--major axis,
not just for the flybys and exchanges, but also the breakups
and triples, and the resultant cross--sections for energy transfer
and recoil velocity distribution. Also of interest are the eccentricity
distributions of the various final binary states. Of particular interest
to us are the properties of the system after it undergoes an
inelastic collision. Simulations of such collisions have been performed
using SPH (Davies \etal 1993, 1994, Goodman \&\ Hernquist 1991,
Sigurdsson \&\ Hernquist 1993). Approximating the collision as a
totally inelastic ``sticky particle'' merger, conserving momentum
but not energy, allows a quick and reasonably accurate way
of determining the properties of the merged systems, in particular
whether they form a single merged star, or if the merged star is
in a binary or even a triple, and if so what the orbital parameters
and center of mass recoil velocity of the system containing the
merged star is. An analysis of these properties is deferred to
a second paper (in preparation).

It is clear that binary--binary interactions are significant for
producing stellar exotica through collisions in globular cluster cores.
Compared to binary--single interactions, the rates inferred suggest
a modest global binary fraction in the cores of the denser clusters,
in accord with previous estimates, with $f_b({\rm all}\  a_i) \sim 0.2$
and $f_b(a_i) \sim 0.05$ per decade in $a_i$, while in the low
density clusters the blue straggler population is consistent
with a somewhat higher binary population, with perhaps $ > 10 \%$
of the turnoff mass main--sequence stars in the core being in
binaries with $a_i \sim 100 AU$. Binary--binary collisions most likely
dominate binary--single collisions in many low density clusters as
suggested by Leonard (1989) and may account for a significant fraction
of the blue stragglers observed.

\section*{ACKNOWLEDGEMENTS}
We thank Dr. Sverre Aarseth for providing some of his integration subroutines.
Research supported in part by
funding provided via the 1995 RGO/IoA/MRAO Summer Student Vacation
Course under the auspices of the RGO and the IoA
and Hampshire LEA. MBD gratefully 
acknowledges the support of the Royal Society through a URF. SS 
thanks the PPARC for support.

\end{document}